\documentclass[aps,prx, twocolumn,amsmath,amssymb,longbibliography,superscriptaddress]{revtex4-2}
\usepackage{amsmath,amssymb,amsfonts,bm}
\usepackage{graphicx}
\usepackage{xcolor}
\usepackage{bbold}
\usepackage{graphicx}
\usepackage{dcolumn}
\usepackage{epstopdf}
\usepackage{epsfig}
\usepackage[caption=false]{subfig}
\usepackage{url}
\usepackage{hyperref}
\usepackage{verbatim}
\usepackage[export]{adjustbox}
\usepackage{scalerel}
\usepackage{braket}

\usepackage{ulem}
\usepackage{enumitem}

\usepackage{amsthm}

\usepackage{lineno}

\newcommand{\frmt}{K_\mathrm{RMT}}

\def\ttau{{\sf t}}

\newcommand{\epsiloneff}{\epsilon_{\mathrm{eff}}}

\newcommand{\Leff}{L_{\mathrm{eff}}}
\newcommand{\teff}{t_{\mathrm{eff}}}

\usepackage{float}
\newcommand{\Tr}{\operatorname{Tr}}

\newcommand{\sigv}{\vec{\sigma}}

\newcommand{\be}{\begin{equation}}
\newcommand{\ee}{\end{equation}}
\newcommand{\ba}{\begin{aligned}}
\newcommand{\ea}{\end{aligned}}

\newcommand{\bmult}{\begin{multline}}
\newcommand{\emult}{\end{multline}}

\newcommand{\kb}{K_\mathrm{F}}
\newcommand{\kc}{K_\mathrm{TI}}
\newcommand{\kd}{K_\mathrm{TIF}}

\newcommand{\kappab}{\kappa_\mathrm{F}}
\newcommand{\kappac}{\kappa_\mathrm{TI}}
\newcommand{\kappad}{\kappa_\mathrm{TIF}}

\def\tth{t_{\rm Th}}
\def\Lth{L_{\rm Th}}

\def\theis{t_{\rm Heis}}

\def\NNTh{\mathcal{N}_{\rm Th}}

\def\epsiloneff{\epsilon_{\rm eff}}

\def\thei{t_{\rm Hei}}

\def\bv{{\bf b}}

\def\TT{w}
\def\VV{v}
\def\NN{\mathcal{N}}

\def\NNeff{\NN_{\rm eff}}
\def\KRMT{K_{\mbox{\tiny RMT}}}

\graphicspath{{Figures/}{Figures/Clean/}}

\newcommand{\average}[1]{\langle #1 \rangle}

\def\Npb{N_{\rm pb}}

\newcommand{\vecb}[1]{{\bf #1}}
\def\ZPotts{Z_{\rm Potts}}

\begin{document}

\newcommand{\titleinfo}{Many-Body Quantum Chaos and Space-time Translational Invariance}

\title{Many-Body Quantum Chaos and Space-time Translational Invariance}  
\author{Amos Chan}
\affiliation{Princeton Center for Theoretical Science, Princeton University, Princeton NJ 08544, USA}

\author{Saumya  Shivam}
\affiliation{Department of Physics, Princeton University, Princeton, New Jersey 08544, USA}

\author{David A. Huse}
\affiliation{Department of Physics, Princeton University, Princeton, New Jersey 08544, USA}

\author{Andrea De Luca}
\affiliation{Laboratoire de Physique Th\'eorique et Mod\'elisation,
CY Cergy Paris Universit\'e, \\
\hphantom{$^\dag$}~CNRS, F-95302 Cergy-Pontoise, France}    

\date{\today}

\begin{abstract}
We study the consequences of having translational invariance in space and in time in many-body quantum chaotic systems. 
We consider an ensemble of  random quantum circuits, composed of single-site random unitaries and nearest neighbour couplings, as a minimal model of translational invariant many-body quantum chaotic systems. 
We evaluate the spectral form factor (SFF)  as a sum over many-body Feynman diagrams, which simplifies in the limit of large local Hilbert space dimension $q$.
At sufficiently large $t$, diagrams corresponding to rigid translations dominate,
reproducing the chaotic behavior of random matrix theory (RMT).
At finite $t$, we show that translational invariance introduces additional mechanisms via two novel Feynman diagrams, known as the \textit{crossed} and \textit{deranged diagrams},  which delay the emergence of RMT.
Our analytics suggests the existence of exact scaling forms which describe the approach to RMT behavior in the scaling limit 
where both $t$ and $L$ are large while the ratio between $L$ and $\Lth(t)$, the many-body Thouless length, is fixed. 
We numerically demonstrate, with simulations of two distinct circuit models, that in such a scaling limit, most microscopic details become unimportant, 
and the resulting scaling functions are largely universal, remarkably being only dependent on a few global properties of the system like the spatial dimensionality, and the space-time symmetries.
\end{abstract}

\maketitle

Understanding the chaotic properties of quantum systems is a notoriously hard problem. 
A fruitful direction has been opened by the combination of two ingredients: First, fingerprints of an underlying chaotic dynamics are visible in the Hamiltonian spectrum of quantum systems~\cite{bohigas1984characterization}; second,
spectral properties are best discussed in statistical terms~\cite{AltshulerShklovskii}. 
This approach eliminates dependence on the microscopic details of the studied systems and brings out the universal characteristics of an ensemble of statistically similar Hamiltonians,
which are captured by the random matrix theory (RMT) contrained only by symmetries~\cite{Brody, Mehta}.
RMT provides a prototype of thermalising dynamics for which the eigenstate thermalisation hypothesis~\cite{Deutsch, Srednicki, Deutsch_2018} is confirmed~\cite{Rigol2008}. 

\begin{figure}[ht]
	\includegraphics[width=0.98\columnwidth
	]{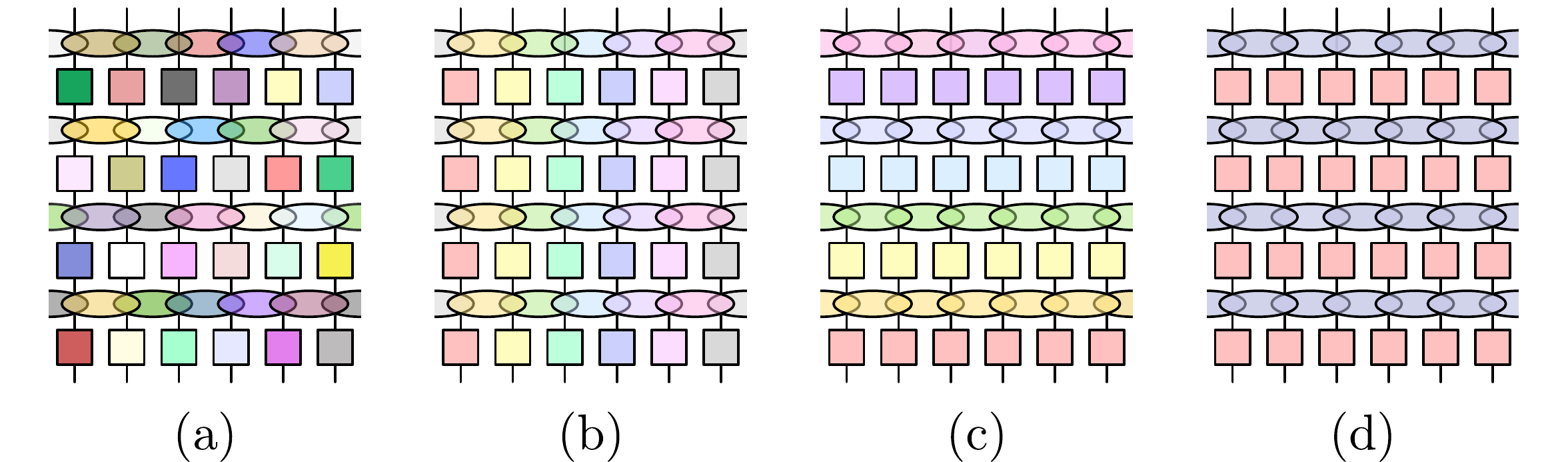}
	\caption{Illustrations of the different types of RUC for the RPM: 
	(a) Temporally and spatially random RPM;
	(b) Floquet (and spatially random) RPM;
	(c)  TI (and temporally random) RPM,
and	(d) TI Floquet RPM.
For each case, gates of the same colors are identical. 
	}\label{fig:model_rpm}
\end{figure}

However, RMT fails to reproduce the local structure of interactions of many-body quantum systems which results in a complex geometry and correlation in the Fock space~\cite{basko2006metal, de2013ergodicity, PhysRevLett.113.046806, PhysRevB.96.201114, PhysRevB.101.134202, PhysRevB.99.104206}. 
For this reason, random unitary circuits (RUC) have been proposed as toy models which utilize RMT while incorporating a notion of locality and dimensionality. 
In the simplest formulation, time evolution of RUC is performed by acting with randomly generated unitary gates on pairs of nearest neighbours in a spin lattice~(Fig.~\ref{fig:model_rpm}a)~\cite{Nahum2017, vonKeyserlingk2017}. 
These models have proven fruitful in developing a unifying picture of the out-of-equilibrium dynamics of generic many-body systems with predictions for the entanglement growth~\cite{Nahum2017, PhysRevB.99.174205, Li_2018, Skinner_2019, Li_2019, chan2019, Gullans_2020, altman2019, Jian_2020, Zabalo_2020}, and the out-of-time-ordered correlators~\cite{Nahum2017a, vonKeyserlingk2017, vonKeyserlingk2017a, Huse2017}. 
More recently, Floquet random unitary circuits (FRUC) have been introduced by applying repeatedly the same set of random gates~(Fig.~\ref{fig:model_rpm}b)~\cite{cdc1, cdc2, cdc3, friedman2019, cdclyap, moudgalya2021}. FRUC have given access to the study of non-trivial spectral properties in extended many-body systems. In particular, for the spectral form factor (SFF) \cite{haake, cdc1, cdc2, friedman2019, cdclyap, moudgalya2021, bertini2018exact, bertini2021random, flack2020statistics, Prosen,  Cotler_2017, complexity2017, Saad2019semiclassical, Gharibyan_2018, li2021spectral, winer2021hydrodynamic, winer2021spontaneous, ZollerSFF2020, ZollerSFF2021, prosen_sff_noise_2021},
\begin{equation}
\label{eq:Kdef}
    K(t, L) \equiv \average{\Tr[W(t)] \Tr[W^\dag(t)]}
\end{equation}
where $W(t)$ is the time evolution operator for time $t$, $L$ is the system size and $\average{\ldots}$ indicates the ensemble average, 
it has been argued that the RMT behavior is recovered only for $t > \tth(L)$, with $\tth(L)$ the SFF Thouless time. $\tth(L)$ is an intrinsic time scale which generally grows unbounded with the system size $L$ (with the exception of the dual-unitary circuits~\cite{Akila_2016, bertini2018exact, bertini2021random, flack2020statistics}). 
Its origin traces back to the existence of domain walls separating growing chaotic subregions~\cite{cdc2, garratt2020manybody, garratt2020MBL}.

In this article, we consider the effect of 
translational invariance in space and time on the SFF. 
We introduce a spatially translational invariant (TI) version of the \textit{random phase model} (RPM) \cite{cdc2} on a $d$--dimensional lattice of length $L$, which can also be time-periodic (Floquet) or not. We will refer to the four setups resulting from the combination of TI and time-periodicity as cases (a), (b), (c) and (d) as illustrated in Fig.~\ref{fig:model_rpm}. 
We show that the SFF is exactly computable in the limit of large local Hilbert space dimension $q$ via a diagrammatic expansion made up of contractions between $\Tr[W(t)]$ and $\Tr[W^\dag(t)]$ (respectively top and bottom layer in Fig.~\ref{fig:summarydiagrams}b) of \eqref{eq:Kdef}.
Before providing the explicit derivation, we outline the main results. 
At large $t \gg \tth(L)$ (but still with $t \ll \theis(L)$, the Heisenberg time which is exponentially large in the system size), 
only \textit{ladder diagrams}, corresponding to rigid translations of the top layer $\Tr[W(t)]$ w.r.t. the bottom layer $\Tr[W^\dag(t)]$, contribute (see Fig.~\ref{fig:summarydiagrams} top). 
This reproduces exactly the RMT predictions, i.e. $K(t, L) \sim \KRMT(t, L)$,
with
\begin{equation}
\label{eq:KRMT}
    \KRMT(t, L) \equiv  
    \begin{cases}
      1 \; , & \text{w/o symm. -- (a)} \\
      t \; , & \text{Floquet -- (b)} \\
      L^d \; ,& \text{TI -- (c)}\\
      t L^d \;, & \text{TI + Floq. -- (d)} 
    \end{cases} \;. 
\end{equation}
The first two lines are standard and result from replacing the time evolution with a random matrix drawn from the circular unitary ensemble (CUE), either re-drawn at every time step (a), or repeated in time (b)~\cite{Mehta}. The remaining lines of Eq.~\eqref{eq:KRMT} can be understood observing that TI on a square lattice leads to $L^d$ momentum sectors, modeled as independent unitary blocks (still drawn from the CUE). This justifies the factors of $L^d$ 
for cases (c) and (d).

\begin{figure*}
\begin{minipage}[t]{0.31\textwidth}
\includegraphics[width=1.1\linewidth,keepaspectratio=true]{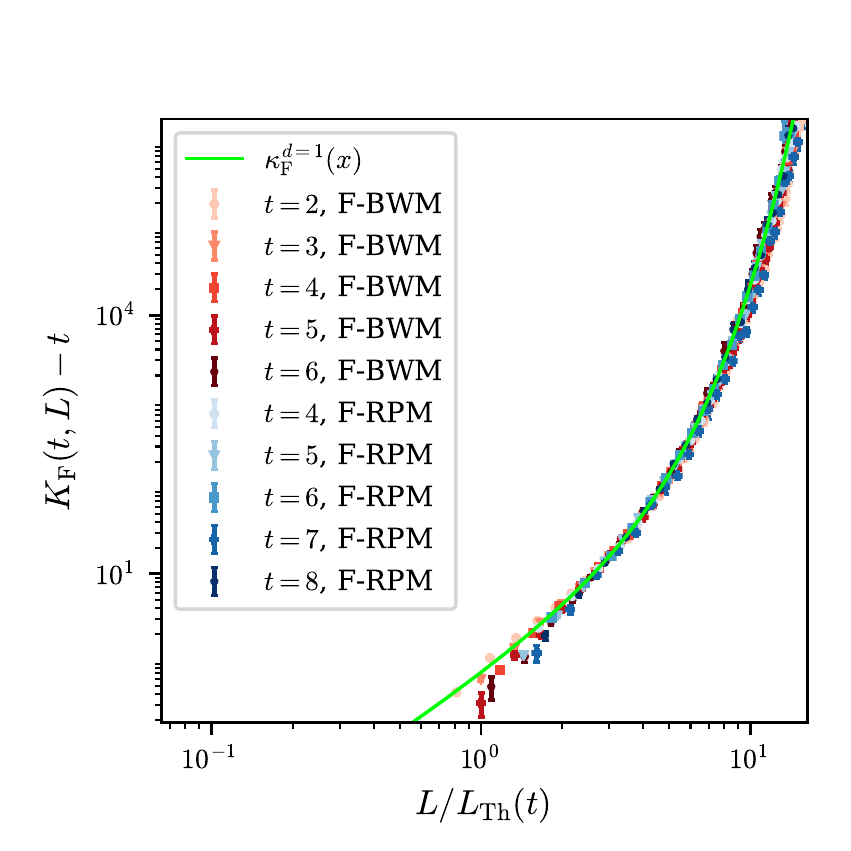}
\end{minipage}
\hspace*{\fill} 
\begin{minipage}[t]{0.31\textwidth}
\includegraphics[width=1.1\linewidth,keepaspectratio=true]{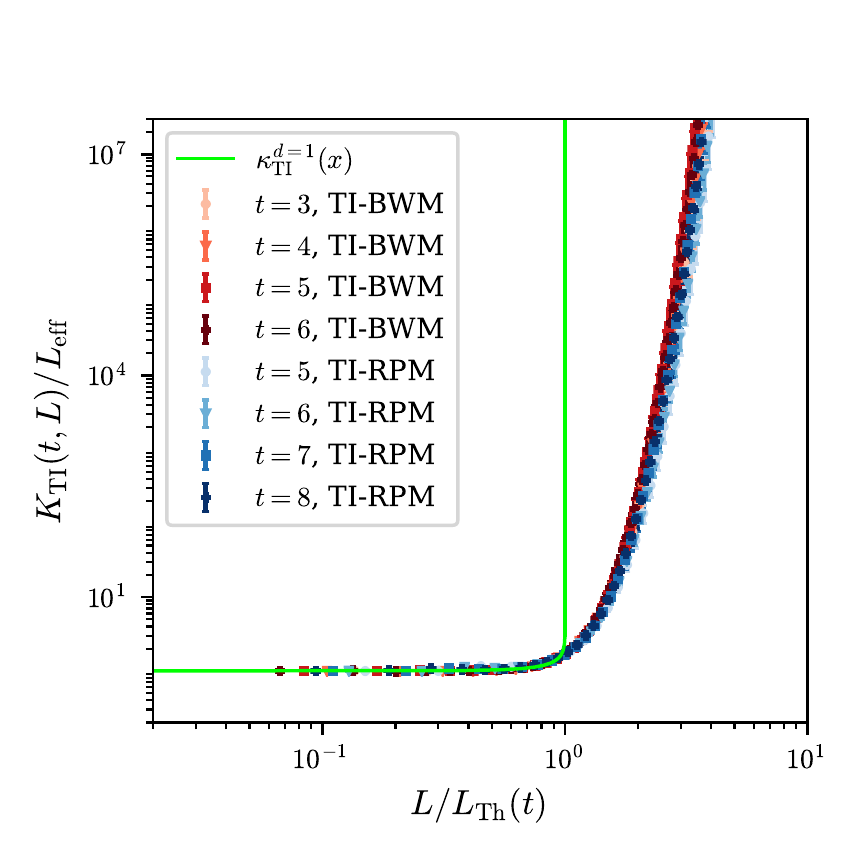}
\end{minipage}
\hspace*{\fill} 
\begin{minipage}[t]{0.31\textwidth}
\includegraphics[width=1.1\linewidth,keepaspectratio=true]{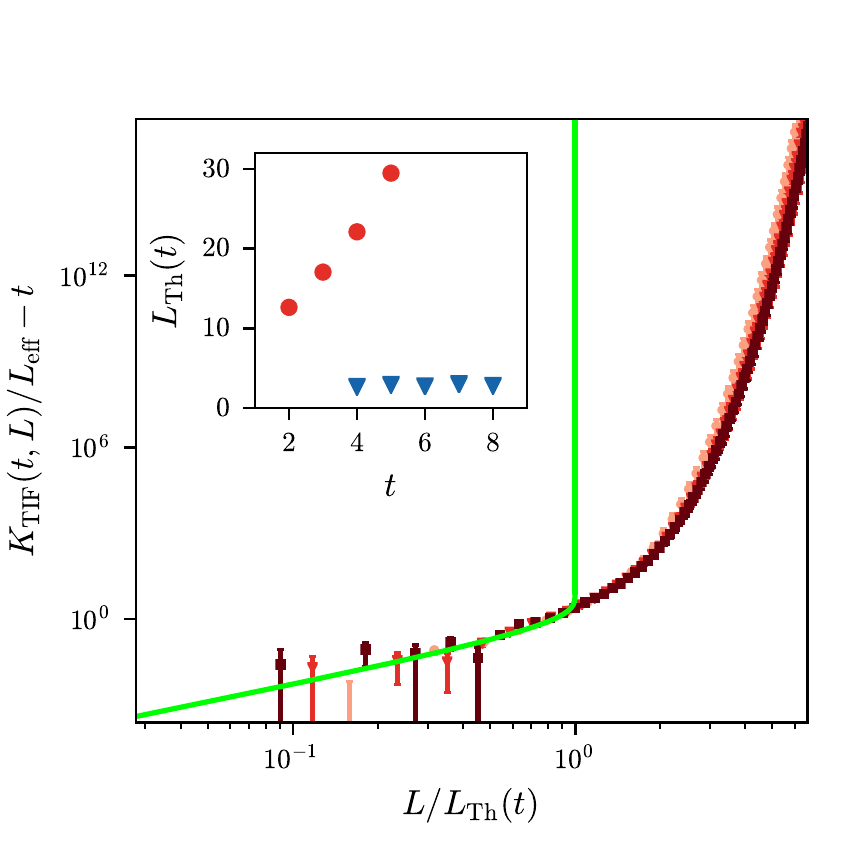}
\end{minipage}
\caption{
Scaled SFF vs. $x=L/ \Lth$ at finite $q$ for Floquet RPM and BWM (left), TI RPM and BWM (middle), and TI Floquet BWM (right); in all panels the RPM has $q=3$, while the BWM has $q=2$ in the left and center panels, and $q=3$ in the right panel. 
The infinite-$q$ scaling forms are plotted in green. 
The main panel of the right plot shows $t =2 ,3,4$ from light to dark red. In the inset, $\Lth$ is plotted against $t$ for TIF BWM (red) and TIF RPM (blue).  The TIF RPM is not shown in the main panel because the apparent $\Lth(t)$ is too small (as seen in the inset) to be reliably estimated~\cite{Note1}.
    } \label{fig:collapse}
\end{figure*}

Additionally, we characterise the corrections to RMT at large but finite $t \gtrsim \tth(L)$. Our diagrammatic calculations at infinite $q$ indicate the existence of scaling forms in the scaling limit where both time $t$ and the system size $L$ are large but the ratio  $L/\Lth(t)$ is kept fixed. 
Here, $\Lth(t)$ denotes the \textit{Thouless length}, defined as the inverse function of the aforementioned Thouless time, i.e. $\Lth(\tth(L)) = L$.
Specifically, we obtain for the relevant cases (b), (c) and (d), the scaling forms 
\begin{align} 
    & \lim_{\substack{L, t \to\infty\\ L/\Lth(t) = x}} \kb(t, L) - t 
    = \kappab(x) \;,   \nonumber 
    \\
    & \lim_{\substack{L,t \to \infty\\ L/L_{\rm Th}(t) = x}} L^{-1} \kc
    =  \kappac (x) \;, \label{eq:sfTI}
    \\
& \lim_{\substack{L, t \to\infty\\ L/\Lth(t) = x}}
L^{-1}
\kd
- t
= \kappad(x) \;. \nonumber  
\end{align}
%
for $d=1$, and for general $d$ below in Sec.~\ref{sec:ana}.  
Remarkably, not only the validity of the scaling forms \eqref{eq:sfTI} obtained at $q\to\infty$ is confirmed by our numerics at finite $q$,  but we also have evidence that the value of the scaling functions $\kappa(x)$ in each case is universal, being independent of the microscopic details of the model, which only affect the non-universal $\Lth(t)$. 

The paper is organized as follows.  
In Sec. \ref{sec:model}, we introduce the random circuit models in the preference of space-time translational invariance.
In Sec.~\ref{sec:numerics}, we discuss the numerical simulations at finite $q$ confirming the existence of the scaling limit and its universality.
In Sec.~\ref{sec:ana}, we derive the scaling forms \eqref{eq:sfTI}
using the large-$q$ diagrammatical computations. 
Lastly, in Sec.~\ref{sec:discussion}, we discuss the discrepencies between the finite-$q$ numerics and large-$q$ analytics, and the possible physical mechanism behind them. 
\section{Models \label{sec:model}}
 The random phase model (RPM) is defined by a quantum circuit which is a matrix product 
$
W(t) = \prod_{t'=1}^t w(t') 
$
where  $w(t)= w_2(t) \, w_1(t) $ is a $q^{\NN}\times q^{\NN}$ operator, with $\NN$ the number of sites. The model can be defined on arbitrary lattices, but here we focus on integer lattices $\mathcal{L}$ in $d$-dimension with periodic boundary conditions and length $L$, so that $\mathcal{L} \equiv \mathbb{Z}_L^d$ and $\NN = L^d$.
$w_1(t) = \bigotimes_{\vecb{r} \in \mathcal{L}} u(\vecb{r}, t) $ 
generates transformations at each site, with $q\times q$ unitary matrices $u(\vecb{r}, t)$ chosen from the CUE;
$w_2(t)$ couples neighbouring sites and is diagonal in the basis of site orbitals with matrix elements
$
[w_2(t)]_{a_1, \ldots a_{\NN};a_1, \ldots a_{\NN}} = 
\exp\left[ \imath \sum_{\langle\vecb{r}, \vecb{r}'\rangle} \varphi_{a_{\vecb{r}},a_{\vecb{r}'}}(\vecb{r}, \vecb{r}', t) \right]$, 
where $\langle \vecb{r}, \vecb{r}'\rangle$ are the nearest neighbours in $\mathcal{L}$ and $a_\vecb{r} \in \{1,\ldots, q\}$.
We take each coefficient $\varphi_{a_{\vecb{r}},a_{\vecb{r}'}}(\vecb{r}, \vecb{r}', t)$ to be a Gaussian random variable with mean zero and variance $\epsilon$, 
which effectively controls the coupling between neighbouring spins.

For the temporally and spatially random RPM 
(Fig.~\ref{fig:model_rpm}a), all unitaries $u(\vecb{r}, t)$ and phases $\varphi(\vecb{r}, \vecb{r}',t)$ are drawn independently. Correlation exists only between unitaries in $\Tr[W(t)]$ and their \textit{unique} conjugates in $\Tr[W^\dag(t)]$, which gives $K(t,L)=1$ for all $q$.
In \cite{cdc2}, the Floquet RPM (Fig.~\ref{fig:model_rpm}b) was considered where all gates are drawn independently in space but are constant in $t$. Here, we also consider the TI RPM (Fig.~\ref{fig:model_rpm}c), where the gates are such that $u(\vecb{r}, t) = u(t), \phi(\vecb{r}, \vecb{r}', t) = \phi^{(\mu)}(t)$, whenever $\vecb{r} - \vecb{r}' = \vecb{e}_\mu$, with $\vecb{e}_\mu$ the unit vector in the $\mu$ spatial direction. The Floquet TI RPM (Fig.~\ref{fig:model_rpm}d), where the gates are also constant in time, is arguably the most realistic set-up. 

\section{Numerics \label{sec:numerics}}
To test the universality of the scaling forms in \eqref{eq:sfTI}, we numerically evaluate the SFF for RPM and additionally the \textit{brick wall model} (BWM) defined in the SM~\cite{Note1} at finite $q=2,3$, $t \leq 8$, and large $L$. Because of the computational cost, higher $d$ are currently out of reach and we restrict to $d=1$.  
The comparison between two different models is shown in Fig.~\ref{fig:collapse}, together with the infinite-$q$ analytical predictions shown in green (see (\ref{eq:kc_scaling}, \ref{eq:kb_scaling_d1},  \ref{eq:kappaTIFd1}) below). 
In all cases, we see an excellent collapse among the different models and times, consistent with the scaling form and pointing at the existence of a universal scaling function. 
We stress that the only free parameter in this procedure is the Thouless length $\Lth(t)$, which rescales the horizontal axis for each $t$. In our procedure, we fix it by imposing that the numerical data at different $t$'s all cross at a reference value $x_0 = \tilde{L}/\Lth(t)$ and equal the infinite-$q$ expression (see Appendix~\ref{app:sfftransl} for details). 
For Floquet circuits (Fig. \ref{fig:collapse} left), using \eqref{eq:sfTI}, we obtain an exceptionally good collapse for both models with the analytic infinite-$q$ calculation (see \eqref{eq:kb_scaling_d1} below). 
Note that while the SFF for the Floquet RPM had already been computed in \cite{cdc2}, the universality of the corresponding scaling function in the scaling limit had not been observed before. 
However, for both TI (non-Floquet) and TI Floquet circuits (Fig. \ref{fig:collapse} middle and right), 
the scaling functions which emerge from numerical data are not well described by those computed at infinite $q$. The physical mechanism behind the discrepancies are discussed further in Sec.~\ref{sec:discussion}.

\begin{figure}[ht]
    \centering
    \includegraphics[width=0.95\columnwidth
    ]{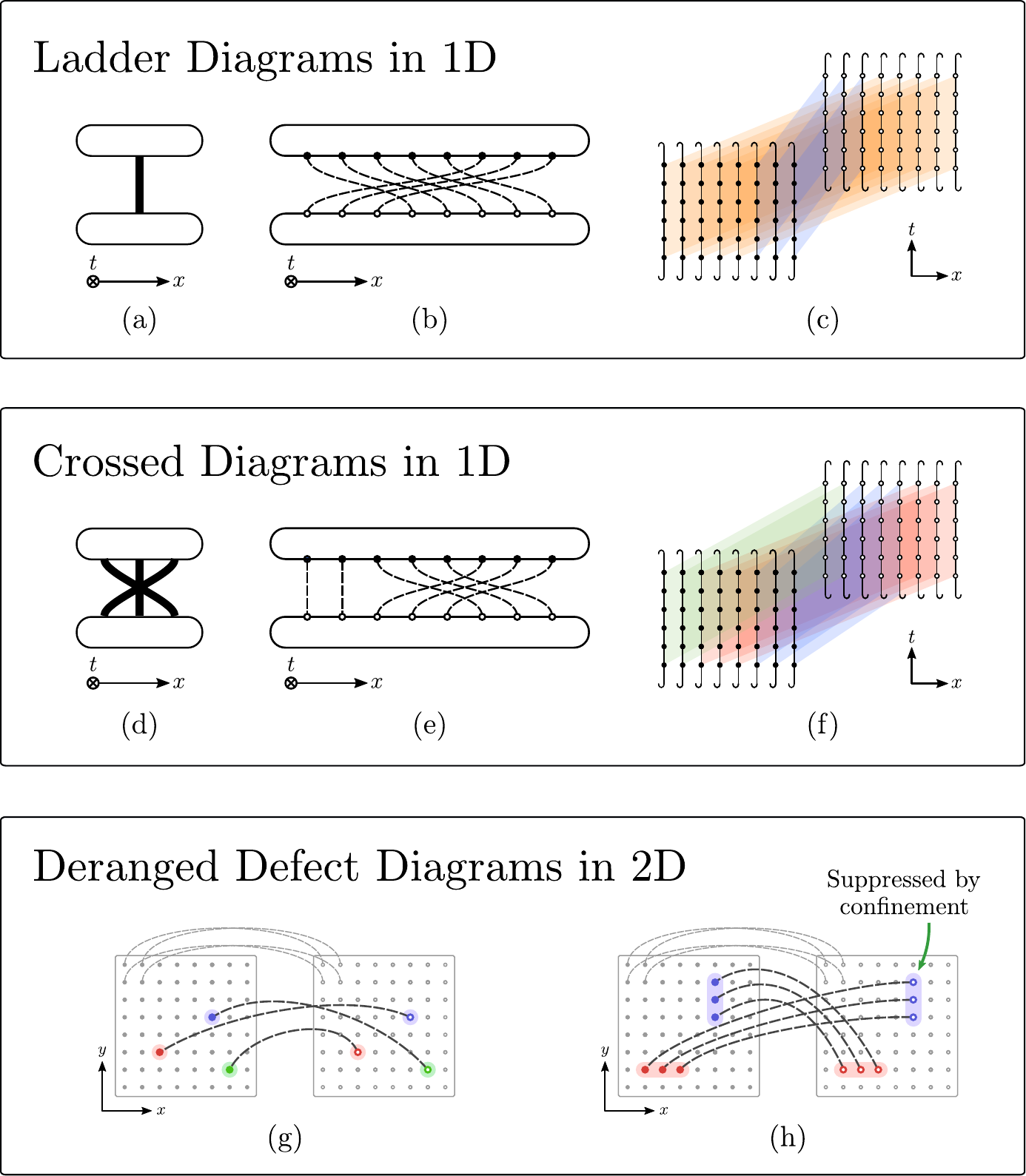}
    \caption{
    (a) A representation of the set of ladder diagrams that appear in SFF of a TI RUC according to the rules presented in \cite{cdc1, cdc2}. 
    The top and bottom layers denote $\Tr[W(t)]$ and $\Tr[W^\dagger(t)]$ respectively. 
    The thick line denote a group of parallel contractions between unitary gates and their conjugates (represented as dashed lines and dots in (b)).  
    This set contains $t$ ladder diagrams which are related to each other by a rigid translations of (say) the top layer,
     and are of the same order in $q$~\cite{cdc1, cdc2}.
    (b) A ladder diagram belonging to the set in (a).
    (c) A many-body Feynman diagram that contributes to the SFF of a TI RUC. 
    The (colored) planes represent a group of parallel contractions.
    (d) 
    The simplest set of crossed diagrams that appear in SFF for TI RUC in 1D.
    (e) A diagram belonging the set represented in (d).
    (f)  A many-body diagram that contributes to the SFF of TI RUC.
    (g) A spatial {\textit{deranged defect diagrams}} that contributes to the SFF for TI RUC in 2D.
    The left and right square represents  $\Tr[W(t)]$ and $\Tr[W^\dagger(t)]$ respectively. 
    Most sites in  $\Tr[W(t)]$  (dots in the left square) are contracted (grey dashed lines) with the same sites in $\Tr[W^\dagger(t)]$ (dots in the right square). 
    We omit these contractions except four of them on the top left corner.
    The sites that do not contract with their counterpart are called \textit{spatial defects} (labelled with colors). 
    (h) Another crossed diagrams that appear in SFF for TI RUC in 2D. 
    The phenomenon of confinement suppresses this diagram with a factor of $e^{-  \epsilon t \partial}$, where $\partial$ is the size of the boundary of the defects (colored in red and blue).
    }
    \label{fig:summarydiagrams}
\end{figure}
\hspace{-0.27cm}

%
\section{Analytics}\label{sec:ana}
Now we sketch the exact large-$q$ analytics of SFF for the TI, Floquet, and TI FLoquet cases systematically. 
The analytics allows us to derive the emergence of the RMT behaviour \eqref{eq:KRMT} in large-$t$, and  the existence of scaling forms \eqref{eq:sfTI} describing the approach towards such emergence.
%
%
%
Remarkably, as demonstrated in Sec.~\ref{sec:numerics} the scaling forms are largely universal and depend only on $q$, spatial dimensionality, and the space-time symmetries.
The dependence of universality classes in the spatial dimensionality was first observed in \cite{cdc2} for Floquet systems and is even more striking for TI ones: In $d=1$, corrections are controlled by \textit{crossed diagrams} where sub-intervals in the top layer are rigidly contracted with those in the bottom layer~(Fig.~\ref{fig:summarydiagrams} middle); 
instead, in $d\geq 2$, corrections are generated by {\textit{deranged defect diagrams}}, where confinement forces excitations (on top of ladder diagrams) to be dilute~(Fig.~\ref{fig:summarydiagrams} bottom).
\subsection{Translational invariant case}
Here we compute the SFF of the TI--RPM in the limit $q\to\infty$.
To compute the SFF, we perform the ensemble averages in two steps: (i) ensemble average over the CUE-s $u(\vecb{r}, t')$; (ii) ensemble averages over the random phases. 
Within a fixed time slice $t'$, there are $\NN$ copies of $u(\vecb{r}, t')$ / $u^{\dag}(\vecb{r}',t')$ on the top/bottom layer. Following \cite{cdc1} 
(see also \footnote{See supplementary material at [url] for analytical computations of SFF for the TI, Floquet and TIF RPM in one- and higher-dimensional cases; for analytics of SFF for circuits with generalised unit cell; for definition of the brick wall model; and for numerical methods and results on SFF and Thouless time $\tth$.}), 
at the leading order in large $q$~\footnote{Throughout the manuscript, we will always take the order of limits where the limit of large $q$ is taken before the limit of large $t$ and $L$ are taken.}, the ensemble average is expressed as a sum over permutations, $\sigma \in S(\NN)$, pairing $u(\vecb{r}, t')$ with  $u^{\dag}(\sigma(\vecb{r}),t')$. 
Additionally, at leading order in large $q$, one is forced to take the same permutation on all $t$ time slices, i.e. SFF is a single sum over $\sigma \in S(\NN)$~\cite{Note1}. 
We now turn to the average over the random phases and we will see that it is natural to interpret it as a \textit{cost function} associated to each $\sigma$. Expanding the orbital sum from phases at time slice $t'$, 
we have
$
    \sum_{a_1,\ldots, a_\NN} \exp 
    \big[
    \imath \sum_{\langle\vecb{r}, \vecb{r}'\rangle} \varphi_{a_{\vecb{r}} ,a_{\vecb{r}'} }
    (t')  - \varphi_{
    a_{\sigma(\vecb{r})} ,a_{\sigma(\vecb{r})} 
    } (t') 
    \big]
$ where the sum is over nearest neighbor pair of sites.
We see that, in large $q$, cancellations of the phases are only possible whenever 
 $\sigma$ maps nearest neighbour sites onto nearest neighbours (preserving the orientation). Using that $\langle e^{\imath \varphi} \rangle = e^{-\epsilon/2}$ and that all time slices contribute equally, we arrive at the expression
\begin{equation}
\label{eq:KTIRPM}
\kc
=
\sum_{\sigma \in \mathcal{S}_{\NN}} e^{-\epsilon t 
 (d \NN - \Npb(\sigma))}
\end{equation}
where $\Npb(\sigma) = \# \{ \vecb{r}, \mu \; |\; \sigma(\vecb{r} + \vecb{e}_\mu) - \sigma(\vecb{r}) = \vecb{e}_\mu\}$ is the number of \textit{preserved bonds} in any direction by the permutation $\sigma$ ($\#$ denotes the cardinality of a set). 
The sum in Eq.~\eqref{eq:KTIRPM} can be reorganized by grouping all $\sigma \in S_\NN$ with the same $N_{\rm pb}(\sigma)$. 
We observe that $\Npb(\sigma) = \Npb(\tau \sigma)$ for all translations $\tau \in S_\NN$. Since the subgroup of translations is isomorphic to the lattice itself $\mathcal{L}$, we arrive at
\be
\label{eq:sffmulti}
\kc
 =
 \NN
\sum_{n=0}^{d \NN}  A_d(\NN, n) e^{- n \epsilon t} \;,
\ee
where $A_d(\NN, n) = \# \{ \sigma \in S_\NN / \mathcal{L} \; |\; \Npb(\sigma) = d \NN - n\}$ and $n = d \NN - \Npb(\sigma)$ is the number of \textit{broken bonds}. 
Computing exactly the $A_d(\NN, n)$ poses a non-trivial combinatorial problem, which nevertheless simplifies in $d = 1$ or for large $\NN$, as we show below. 
However, it is easy to see that $A_d(\NN, n = 0) = 1$, corresponding to the identity equivalence class. Therefore in the limit $t \to \infty$, we recover the RMT result for this case in Eq.~\eqref{eq:KRMT}. 
Generalising this construction, one can see that $K$ converges to the dimension of the group of spatial symmetries. 
As example, for a two-dimensional TI circuit on a square lattice with rotational symmetry by angles of $\pi/2$, $\kc = \mathcal{N} = 4$ in the limit of large $t$. 

For $d = 1$, $\sigma \in S_L / \mathbb{Z}_L$ can be represented as cyclic permutations of $L$ elements: Represent $\sigma \in S_L / \mathbb{Z}_L$ as $\sigma' \equiv (\sigma(1),\sigma(2),\dots)$ and define an associated cyclic permutation with cycle $\sigma'$.
Then, $\sigma$ with a fixed number of broken bonds $n$ can be obtained as follows: We first partition $\mathcal{L} \equiv \mathbb{Z}_L = I_1\cup I_2\ldots I_n$ into $n$ adjacent non-empty intervals. 
Then we take any cyclic permutation $\tilde{\sigma}$ of $n$ elements such that $\tilde{\sigma}(i) \neq \tilde{\sigma}(i+1) + 1 \pmod n$
and rigidly map $I_i \to I_{\tilde{\sigma}(i)}$. 
Clearly, the resulting mapping breaks exactly $n$ bonds and all possible mappings can be uniquely constructed in this way.
As an example, see Fig.~\ref{fig:summarydiagrams}e where $L=8, n=3$ and $\tilde{\sigma} = (132)$.
This leads to
\begin{equation}\label{eq:case_c_1d_identity}
    A_1(L, n) = \binom{L}{n} a_n
\end{equation}
where the binomial factor counts the partitionings of $\mathcal{L}$ and $a_n$ the possible $\tilde \sigma$. Although an explicit expression for the $a_n$ is not available, they correspond to a well-studied sequence whose exponential generating function is known~\cite{oeisA000757, stanley2011enumerative}. 
For both $L, t$ large with fixed $x = L/\Lth(t)<1$, $\Lth(t) = e^{\epsilon t}$, we can take $\binom{L}{n} \sim L^n/n!$ using the dominated convergence theorem~\cite{Bar1995} and sum over $n$ to obtain the scaling form
\begin{equation} \label{eq:kc_scaling}
    \lim_{\substack{L,t \to \infty\\ L/L_{\rm Th}(t) = x}} L^{-1} \kc^{d=1} = e^{-x} (1 - \ln (1 - x)) \equiv \kappac^{d=1}(x) \;.
\end{equation}
$\Lth(t)$ denotes the Thouless length. 
Note also that throughout the article, the limit of large $q$ is always taken before the limits of large $t$ and $L$.

For $d > 1$, evaluating the multiplicities $A_d(\NN,n)$ is a much harder task as they result from the interplay between permutations and the geometry of $\mathcal{L}$. Nevertheless, the problem simplifies in the limit of large $L$ at fixed $n$, as it corresponds to a dilute regime where a fixed number of bonds is broken in a very large system. 
Consider first a transposition which exchanges two sites. This will generally break $4d$ bonds ($2d$ neighbours for each site), and therefore $A_{d>1}(\NN, 4d) \sim \NN^2/2$. 
More generally, we first pick the positions of $k$ well-separated spatial defects and then we consider the possible ways of permuting them without leaving fixed points, so that precisely $n = 2d k$  bonds are broken (Fig.~\ref{fig:summarydiagrams}g). These {\textit{deranged defect diagrams}} lead to the asymptotic expansion
\begin{equation}
\label{eq:AlargeLdgt1}
    A_{d}(\NN, n = 2 d k) \stackrel{\NN \to \infty}{=} \frac{\NN^k}{k!} d_k \;, \quad d > 1
\end{equation}
where $d_k$ are the number of \textit{derangements}~\cite{graham1989concrete, stanley2011enumerative} (i.e. permutation with no fixed points) of $k$ elements~\cite{oeisA000166}. 
The error we make in \eqref{eq:AlargeLdgt1} is related to situations where the defects are close to one another thus breaking less bonds, but these are sub-leading in $\NN$. Once again, we consider the limit $t, L \to \infty$ at fixed $x = \NN / \NN_{\rm Th}(t)$, with $\NN_{\rm Th}(t) = e^{2 d \epsilon t}$ the \textit{Thouless volume},
and obtain
\begin{equation} \label{eq:kc_scaling_dgt1}
    \lim_{\substack{L,t \to \infty\\ \NN/\NN_{\rm Th}(t) = x}} \NN^{-1} \kc^{d>1} = \frac{e^{-x}}{1-x} \equiv \kappac^{d>1}(x)
\end{equation}
We stress that the apparent difference between the scaling functions \eqref{eq:kc_scaling} and \eqref{eq:kc_scaling_dgt1} has a fundamental origin: In $d=1$, extended intervals can be rigidly exchanged paying a cost only at their boundary; instead, in $d > 1$, exchanging two extended domains has a cost which grows with their boundary, i.e. extended defects are suppressed by \textit{confinement}. 
Therefore, the leading contribution at large $L$ and $t$ is given by well-separated single-site excitations. 
\subsection{Floquet case}
Before analysing the effect of combining time periodicity and translation invariance, we review the calculation of $\kb(t,L)$ for the Floquet RPM (Fig.~\ref{fig:model_rpm}a)~\cite{cdc2}. Here the single-site unitary gates are constant in time but are random in space $u(\vecb{r}, t) = u(\vecb{r})$. Thus, 
in the diagrammatic expansion of the SFF, we can choose any permutation $\sigma \in S_t$ to pair the $t$--copies of $u(\vecb{r})$ in the top layer with those of $u^\dag(\vecb{r})$ in the bottom one. In the limit of large $q$, only time translations contribute, i.e. $\sigma_\ttau(k) = k + \ttau \pmod{t}$ with $\ttau = 0,\ldots, t-1$~\cite{cdc2}. Therefore, we get a many-body diagram by choosing a configuration $\ttau(\vecb{r})$ (color) for each site $\vecb{r} \in \mathcal{L}$. After averaging over the random phases, it was shown that $\kb(t,L) = \ZPotts$, with $\ZPotts$ the partition function of a $t$-state Potts model with a Boltzmann weight across all bonds $\mathcal{W}_{\ttau, \ttau'} = e^{-\epsilon t (1-\delta_{\ttau, \ttau'})}$. 
At large $t$, the partition function is dominated by the $t$ ferromagnetic groundstates where all sites have the same color, leading to the RMT prediction $K_{\rm F} \sim t$. 
As $t$ approaches $\tth$ from above, excitations from the $t$ ferromagnetic groundstates -- the lowest-lying excitation being the domain wall states \cite{cdc2, garratt2020dw} -- become important.
To access such corrections, in 1D, one makes use of the transfer matrix to write $\ZPotts = \Tr[\mathcal{W}^L]$. Computing the spectrum of $\mathcal{W}$ and evaluating $\ZPotts$ in the scaling limit, this leads to the scaling form~\cite{Note1}
\begin{equation}
\label{eq:kb_scaling_d1}
\lim_{\substack{L, t \to\infty\\ L/\Lth(t) = x}} \kb^{d=1}- t = e^x - x - 1 \equiv \kappab^{d=1}(x)
\end{equation}
with $x = L / \Lth(t)$ and $\Lth(t) = e^{\epsilon t}/t$. 
In higher dimension,  $\ZPotts$ cannot be easily computed for finite $L$ and $t$. Nevertheless, in the scaling limit where $L,t$ are both large, corrections to the RMT SFF are associated with diluted excitations where the color is changed with respect to the ground state's one. 
The position of the excitation can be chosen in $\sim\NN^n / n!$ ways and each of them can be assigned any of the $t-1$ remaining color, with a cost $e^{-2 n \epsilon d t}$. As a consequence, setting $x=\NN/\NN_{\rm Th}(t)$ and $\NN_{\rm Th}(t) = e^{2 d \epsilon  t}/t$,
\begin{equation}
\label{eq:kb_scaling_dgt1}
\lim_{\substack{L, t \to\infty\\ \NN/\NN_{\rm Th}(t) = x}} t^{-1} \kb^{d>1}= e^x \equiv \kappab^{d>1}(x) \;.
\end{equation}
Intriguingly, note that, in contrast with Eqs.~\eqref{eq:kc_scaling} and \eqref{eq:kc_scaling_dgt1} which are divergent for any $x\geq 1$, Eqs.~\eqref{eq:kb_scaling_d1} and \eqref{eq:kb_scaling_dgt1} remain always smooth for finite $x$. This can be understood observing that infinite-$q$ TI systems are mapped onto stat-mech models with non-local interactions, so that the scaling function is associated with the exchange of distant domains ($d = 1$) or defects ($d > 1$); on the contrary for Floquet systems, the resulting Potts model has purely local interactions.

\subsection{Translational invariant Floquet case}
%
We can now turn to the TI Floquet case. The same matrix CUE matrix $u$ and $u^\dag$ appear $t \NN$ times respectively in the top and bottom layer.  However, at large $q$ only the subgroup $S_{\NN} \times \mathbb{Z}_t^\NN \subset S_{t \NN}$, corresponding to arbitrary spatial permutations $\sigma$ and time translations $\ttau(\vecb{r})$ at each site. 
In $d=1$, as explained in the TI case, the permutation $\sigma$ corresponds to crossed diagrams, where spatial intervals in the top layer $\Tr[W(t)]$ are mapped onto intervals  in the bottom one $\Tr[W^\dag(t)]$ (e.g. Fig.~\ref{fig:summarydiagrams}e).
Then, the cost associated to the average over the phases depends on the choices of $\ttau(\vecb{r})$: within the same interval, the cost is given by the Boltzmann weights $\mathcal{W}$ as in the Floquet case; instead, between different intervals, the cost is always $e^{- \epsilon t}$ irrespectively of the choice of $\ttau$'s at the interface. 
To account for the resulting combinatorics, it is useful to introduce the partition function $Z(\omega) = \Tr((\mathcal{W} + \omega \mathcal{R})^L) = \sum_{n} \omega^n Z_n(t,L)$, where $\mathcal{R}_{\ttau,\ttau'} = e^{-\epsilon t}$ is a rank 1 matrix with constant coefficients. 
In words, $n$ counts the number of intervals and  the factors $Z_n(t,L)$ contain the sum over all possible the colors with a $n$ intervals. 
This leads to the explicit formula valid for arbirtrary $t$ and $L$
\begin{equation}
    \kd^{d=1}(t,L) = \sum_{n=0}^\infty a_n Z_n(t, L) \;.
\end{equation}
In the scaling limit, we obtain the behavior~\cite{Note1}
\begin{equation}
\label{eq:kappaTIFd1}
\lim_{\substack{L, t \to\infty\\ L/\Lth(t) = x}}
L^{-1}
\kd^{d=1}
- t
= \ln \left(\frac{e^{-x}}{1 - x}\right) \equiv     \kappad^{d=1}(x) \;.
\end{equation}

In $d>1$, the large-$t$ dominant contribution corresponds to ladder diagrams in space and a ferromagnetic groundstate in the color, leading as expected to $\kd^{d>1}(t, L) \sim t L^d$. Corrections at large $L, t$ are once again obtained by diluted excitations which can have two different origins: derangements as in \eqref{eq:kc_scaling_dgt1}, which are now {\textit{deranged defect diagrams}} in space-time; or color changes as in \eqref{eq:kb_scaling_dgt1}. The two effects combine multiplicatively~\cite{Note1} 
\begin{equation}
\label{eq:kappaTIFdgt1}
\lim_{\substack{L, t \to\infty\\ 
\NN/\NN_{\rm Th}(t) = x}}
(\NN t)^{-1} \kd^{d>1} = 
\frac{1}{1-x}\equiv  \kappad^{d>1}(x) \; ,
\end{equation}
with $\NN_{\rm Th}(t) = e^{2 d \epsilon t}/t $.  

\section{Discussion \label{sec:discussion}}
The numerics shown in Fig.~\ref{fig:collapse} (middle and right) shows a discrepancy between the infinite-$q$ analytics and the finite-$q$ numerics in the presence of space translation invariance. 
There can be different justifications behind this discrepancy: One possibility is that finite $t$ corrections decay very slowly for TI systems. This is qualitatively confirmed in the infinite-$q$ case, inspecting how the limit in \eqref{eq:kc_scaling} is approached increasing $t$~\cite{Note1}. A slow convergence is also to be expected due to the presence of singularities at finite $x$ in (\ref{eq:kc_scaling},  \ref{eq:kappaTIFd1}).
More probably, we have indications that Eq.~\eqref{eq:kb_scaling_d1} is more robust than \eqref{eq:kc_scaling}: By looking at the Floquet/TI RPM with $p$-site unit cell at infinite $q$, we find that the scaling function \eqref{eq:kb_scaling_d1} is independent of $p$ while \eqref{eq:kc_scaling} is not~\cite{Note1}.
Still, it might appear
puzzling that finite-$q$ numerics shows a collapse to a scaling function indepedent of $q$, which is nonetheless not in agreement with the infinite-$q$ analytics. 
To elucidate this aspect, we propose a simple qualitative scenario. 
First, we observe that in the RPM, the two parameters $\epsilon$ and $q$ control respectively the coupling in the space and time directions and the RMT behavior emerges when long-range order is established in both directions. 
In $d=1$, corrections to RMT are then controlled by dilute excitations which break ordered domains, either in space or time, i.e. in the leading order, we have $K(t, L) \sim \KRMT(t,L) + g_0(L/ L_{{\rm Th}, 0}(t) ) + g_1(L/L_{{\rm Th}, 1}(t))$; where the subscript $0$ and $1$ refer respectively to the time and space directions, with the corresponding correlation lengths $L_{{\rm Th}, 0 / 1}(t)$. The functions $g_0(x)$ and $g_1(x)$ tend to zero as $x\to 0$, and are expected to be model independent and only dependent on the symmetries. 
In the large $q$ limit at fixed $L$ and $t$, the coupling in the time direction becomes infinitely strong with  $L_{{\rm Th}, 0}(t) \stackrel{q \to \infty}{\longrightarrow} \infty$
so that only the function $g_1(x)$ survives in the decomposition above. 
For the Floquet RPM, we expect the relevant length scale to be $\Lth(t) \sim L_{{\rm Th}, 1}(t) \ll L_{{\rm Th}, 0}(t)$, so that the scaling function is dominated by spatial domain walls, i.e. $\kappab^{d=1}(x) \sim g_{1, {\rm F}}(x)$.
However, for TI systems, $\Lth(t) \sim L_{{\rm Th}, 0}(t) \ll L_{{\rm Th}, 1}(t)$, and therefore $\kappac^{d=1}(x) \sim g_{0, {\rm TI}}(x)$. 
However, as pointed out above, this scaling function is not accessible if $q \to \infty$ before $L$ and $t$, thus explaining the observed discrepancy between the numerics and the analytics. 
In practice, the universal scaling function $g_{0, {\rm TI}}(x)$ observed in Fig.~\ref{fig:collapse} results from \textit{temporal domain walls} where contractions in different time slices take different permutation values. We stress that, because of unitarity, the SFF cannot diverge exponentially in $t$, therefore temporal domain walls
must contribute \textit{both} positively and negatively, in a distinct contrast to the spatial domain walls discussed in \cite{cdc2, garratt2020dw}. This will be discussed in an upcoming work \cite{upcoming}.

A few additional comments are in order. 
Firstly, it would be beneficial to justify the universality which emerges from our work by means of a well-defined renormalization procedure. The main difficulty in this direction are the lack of locality and positivity of the resulting stat-mech model.
Secondly, it is natural to expect that the scaling regime we identified is also visible in other quantities, as time-dependent correlation functions like out-of-time-ordered correlators.
Thirdly, quasimomentum is conserved in TI lattice systems and affects its spectral properties but does not lead to the transport of an extensive conserved quantity because of Umklapp scattering; it is therefore interesting to explore its interplay with $U(1)$ conserved charges. This will be discussed in an upcoming work~\cite{upcoming}.

\section{Acknowledgements.} 
AC and ADL warmly thank John Chalker for his guidance in related projects.  DAH thanks Grace Sommers and Michael Gullans for a related collaboration.  DAH is supported in part by NSF QLCI grant OMA-2120757.
AC is supported by fellowships from the Croucher foundation and the PCTS at Princeton University. 

\bibliography{biblio.bib}

\onecolumngrid
\newpage 

\appendix
\setcounter{equation}{0}
\setcounter{figure}{0}
\renewcommand{\thetable}{S\arabic{table}}
\renewcommand{\theequation}{S\thesection.\arabic{equation}}
\renewcommand{\thefigure}{S\arabic{figure}}
\setcounter{secnumdepth}{2}

\begin{center}
{\Large Supplementary Material \\ 
\vspace{0.22cm}
\titleinfo
}
\end{center}

In this supplementary material we provide additional details about:
\begin{enumerate}[label=\Alph*)]
    \item SFF for translational invariant RPM
    \begin{itemize}
        \item[1. ] Derivation of Eq.~\ref{eq:KTIRPM}
        \item[2. ] One dimensional case
        \item[3. ] Higher dimensional case         
    \end{itemize}

    \item SFF for Floquet RPM.
    \begin{itemize}
        \item[1. ] One dimensional case
        \item[2. ] Higher dimensional case 
    \end{itemize}

    \item SFF for translational invariant Floquet RPM
    \begin{itemize}
        \item[1. ] One dimensional case
        \item[2. ] Higher dimensional case 
    \end{itemize}
    
    \item Generalised unit cell in large-$q$
    \begin{itemize}
    \item[1. ] SFF for TI RPM with $p$-site translational invariance
    \item[2. ] SFF for Floquet RPM with $p$-discrete-time translational invariance
    \end{itemize}
    
    \item Brick wall model (BWM)
    
    \item Numerical methods and results
    \begin{itemize}
        \item[1. ] SFF for translational invariant circuits
         \item[2. ] SFF for Floquet circuits
        \item[3. ] SFF for translational invariant Floquet circuits
        \item[4. ] Comparison between scaling forms and finite-$t$, finite-$L$, and infinite-$q$ solutions
    \end{itemize}
\end{enumerate}


\section{SFF for translational invariant RPM}
\subsection{Derivation of Eq.~\ref{eq:KTIRPM}}
To derive Eq.~\ref{eq:KTIRPM}, we compute the ensemble average of the TI-RPM in two steps: the ensemble average  of (i) the unitaries drawn from the CUE in $w_1(t')$, and (ii) the random phases in $w_2(t')$.
To carry out (i), we recall that the ensemble averages of unitaries $u_{i,j}$ drawn the CUE can be evaluated as
$
\average{
u_{i_1 j_1}  u_{i_2 j_2} \dots u_{i_\ell j_\ell}
u^*_{i'_1, j'_1}
u^*_{i'_2, j'_2}
\dots u^*_{i'_\ell, j'_\ell}
}
=
\sum_{\sigma, \tau \in S_\ell} 
\mathrm{Wg}(\sigma^{-1} \tau , q)
\prod_{k=1}^\ell \delta_{i_{\sigma(k)}, i'_{k}} 
\delta_{j_{\tau(k)}, j'_{k}}
$
where $S_\ell$ is the symmetric group of $\ell$ elements, and $\mathrm{Wg}(\sigma^{-1} \tau, q)$ is the Weingarten function~\cite{Samuel, Mello}. 
$\mathrm{Wg}(\sigma^{-1} \tau, q)$ can be expressed as a polynomial in $1/q$, and is of the leading order in $q$ when $\sigma= \tau$. 
We apply the above formula to average over unitaries in a fixed time slice $t'$.
In the large-$q$ limit, the sum over permutation is dominated by the term with $\sigma = \tau \in S_{\NN}$, where $\NN$ is the number of sites in the lattice. 
Together with all $t$ time slices, the ensemble average is a sum over a vector of permutations, $\sigv = (\sigma_1, \sigma_2, \dots, \sigma_t)$, where $\sigma_i \in S_{\NN}$.
Observe that the average over the random phases can only maintain or decrease the order in $1/q$ of a given contribution. 
This implies that in the large-$q$ limit, the leading contribution must have $\sigma_1= \sigma_2 =\dots = \sigma_t = \sigma \in S_{\NN}$, so that the number of sums over the site orbitals is maximized. 
Next, we perform the ensemble averages of the random phases in (ii).
Expanding the orbital sum from phases at time slice $t'$, we have
\be \label{eq:rp_case_c}
\sum_{\{\vecb{a}\}, \{\vecb{a}'\} =1}^q
\prod_{\alpha, \beta= 1}^q 
\int D\varphi_{\alpha \beta}
\prod_{\vecb{r} }^{\NN}
e^{\imath 
\sum_{\mu}
\left[ 
\varphi_{a_{\vecb{r}} a_{\vecb{r} + \vecb{e}_\mu}}
- \varphi_{a'_{\vecb{r}} a'_{\vecb{r} + \vecb{e}_\mu}} 
\right]
}
\delta_{a_{\vecb{r}}, a'_{\sigma( \vecb{r}) }  } 
 \;, 
\ee
where $D\varphi_{\alpha \beta} =  
\left[
{ \exp\left(
{-\frac{\varphi^2_{\alpha \beta}}{ 2\epsilon }}
\right)
}
/
{\sqrt{2\pi \epsilon}} 
\right]
d \varphi_{\alpha \beta}  $, and where $\{\vecb{a} \} $ and $\{\vecb{a}' \} $ labels the orbital degrees of freedom in $\Tr[W(t)]$ and $\Tr[W^\dag(t)]$ respectively. 
The delta functions in \eqref{eq:rp_case_c} will lead to cancellations of phases in the exponent, when $\sigma(\vecb{r}+ \vecb{e}_\mu)  - \sigma(\vecb{r})  =  \vecb{e}_\mu$. 
In the large-$q$ limit, each remaining phase $\varphi_{a_{\vecb{r}} a_{\vecb{r} + \vecb{e}_\mu}}$ will give rise to a factor of $\langle e^{\imath \varphi} \rangle = e^{-\epsilon/2}$. 
Furthermore, since all time slices take the same contraction $\sigma$, the factors from \eqref{eq:rp_case_c} are raised to the $t$-th power. 
This gives us the expression Eq.~\eqref{eq:KTIRPM}.
 
\subsection{One dimensional case}
Consider first of all a possible partitioning of $\mathcal{L}$ into $n$ intervals $I_1,\ldots,I_n$. We define the set
\begin{equation}
\label{eq:Mndef}
    M_n = \{ \tilde\sigma \in S_n/\mathbb{Z}_n |  \forall i \;, \quad \tilde\sigma(i+1) \neq \sigma(i) + 1 \pmod{n} \} \;.
\end{equation}
As explained in the main text, we denote as $a_n$ its cardinality: $a_n = \# M_n$. The sequence $a_n$ satisfies
\begin{equation}
\label{eq:genfunan}
    \sum_{n} \frac{a_n z^n}{n!} = e^{-z} (1 - \ln (1 - z) )  \;,
\end{equation}
which converges for all $|z|<1$. Therefore, taking $L$ large in Eq.~\eqref{eq:case_c_1d_identity}, we have
\begin{equation}
  \lim_{\substack{L,t \to \infty\\ L/L_{\rm Th}(t) = x}} L^{-1} \kc^{d=1} =
  \sum_{n=0}^\infty \frac{a_n x^n}{n!} 
=e^{-x} (1 - \ln (1 - x)) \equiv \kappac^{d=1}(x)   \;,
\end{equation}
where $x=L/\Lth(t)$ and $\Lth(t) = e^{\epsilon t}$, as given in Eq.~\eqref{eq:kc_scaling} of the main text.

\subsubsection{$x\geq 1$ regime}
According to the analysis in the previous section, we saw that the scaling function $\kappac^{d=1}(x)$ diverges when $x\geq 1$. We can investigate how does it diverge at finite but large $L$. In order to do so, we consider the exact expression for finite $L$ at infinite $q$ and $d = 1$, in Eqs.~\eqref{eq:KTIRPM} and \eqref{eq:case_c_1d_identity}.
\be
\label{eq:sffmultid1}
 \kappac^{d=1}(x; L) \equiv 
\sum_{n=0}^{L}  \binom{L}{n} a_n x^nL^{-n} \;,
\ee
where we fixed $x = L/\Lth$. Since for $x\geq 1$, the expression in \eqref{eq:sffmultid1} diverges at large $L$, its behavior is dominated by the large $n$ expansion of the $a_n$ coefficients which can be deduced from \eqref{eq:genfunan} and reads
\begin{equation}
    \label{eq:largenan}
    a_n \simeq \frac{n!}{e (n+1)}  \equiv \tilde{a}_n
\end{equation}
We can thus split the sum \eqref{eq:sffmultid1} as
\be
\label{eq:sffmultid12}
 \kappac^{d=1}(x; L) \simeq 
\sum_{n=0}^{L_0}  \binom{L}{n} (a_n - \tilde{a}_n) x^nL^{-n}  + \sum_{n=0}^{L}  \binom{L}{n} \tilde{a}_n x^n L^{-n} \;,
\ee
where $L_0$ is large but finite. The first term in \eqref{eq:sffmultid12} converges to the finite contribution
\begin{equation}
\sum_{n=0}^{L_0}  \binom{L}{n} (a_n - \tilde{a}_n) x^nL^{-n} \quad \stackrel{L \to \infty}{\longrightarrow}
\quad  \sum_{n=0}^{L_0}  \frac{(a_n - \tilde{a}_n)x^n}{n!}
\end{equation}
To evaluate the second term in \eqref{eq:sffmultid12}, we use the integral representation of the factorial as
\begin{equation}
    n! = \int_0^\infty dt \; e^{-t} t^n 
\end{equation}
which leads to
\begin{equation}
\sum_{n=0}^{L}  \binom{L}{n} \tilde{a}_n x^n L^{-n} = \frac{L}{e x(L+1)}\int_0^\infty \frac{dt}{t} \; e^{-L t} \left((1 + t x)^{L+1}-1\right)
\end{equation}
For large $L$ and $x>1$, this last integral can be estimated via saddle point at $t^{\ast} = (x-1)/x$, which leads to
\begin{equation}
\kappac^{d=1}(x; L) \propto 
\exp[L ( \ln x - 1 + 1/x)]
\end{equation}
For $x = 1$, the integral is dominated by small $t^{\ast} \propto 1/L$, which leads eventually to
\begin{equation}
    \kappac^{d=1}(x; L) \simeq \frac{\ln L}{2 e} + O(1)
\end{equation}

\subsection{Higher dimensional case}
For higher dimension, we cannot obtain an explicit expression for $K_{\rm TI}^{d>1}$ at finite $\NN$ and $t$. However, we are still interested in the limit of large $L$ and $t$. In this case, once again we are interested in the dilute limit, where a permutation only exchanges a fixed number of sites without leaving fixed points. For the number of derangements $d_n$, we have the exponential generating function
\begin{equation}
    \sum_{n=0}^{\infty} \frac{d_n z^n}{n!} = \frac{e^{-x}}{1-x} \;.
\end{equation}
Then, using \eqref{eq:sffmulti} and \eqref{eq:AlargeLdgt1} in the main text, we arrive at
\begin{equation}
    \lim_{\substack{L, t \to \infty \\ \NN / \NN_{\rm Th}(t) = x}} \NN^{-1} K_{\rm TI}^{d>1}(t, L)
    = \sum_{n = 0}^{\infty} \frac{x^n d_n}{n!} 
     = \frac{e^{-x}}{1-x}
    \equiv \kappac^{d>1}(x) \;,
\end{equation}
where $x = \NN / \NN_{\rm Th}(t)$ and $\NN_{\rm Th}(t) = e^{2 d \epsilon t}$, as given in Eq.~\eqref{eq:kc_scaling_dgt1} of the main text.


\section{SFF for the Floquet RPM}
As reviewed in the main text, the calculation of the SFF in the Floquet RPM~\cite{cdc2} at $q\to\infty$ can always be mapped into the partition function of a corresponding Potts model
\begin{equation}
    \kb(t, L) = \ZPotts \;,
\end{equation}
whose Boltzmann weights are given by the matrix $W_{\ttau, \ttau'} = e^{-\epsilon t ( 1- \delta_{\ttau,\ttau'})}$ and $\ttau,\ttau' = 0,\ldots, t-1$. 
\subsection{One dimensional case}
In 1d, we can simply express the partition function by using the Transfer matrix
\begin{equation}
    \kb^{d=1}(t,L) = \Tr[\mathcal{W}^L] \;.
\end{equation}
The spectrum of the matrix $\mathcal{W}$ contains two eigenvalues $\lambda_+ > \lambda_-$: $\lambda_- = 1 - \exp(-\epsilon t)$ with degeneracy $t-1$ and $\lambda_+ = 1 + e^{-t \epsilon}(t-1)$. Consequently,
\begin{equation}
\label{eq:KFd1lambda}
    \kb^{d=1}(t,L) = (t-1) \lambda_-(t)^L + \lambda_+(t)^L   \;. 
\end{equation}
In this case, setting $\Lth(t) = e^{\epsilon t}/t$ and $x = L/\Lth(t)$, we have
\begin{equation}
\lim_{\substack{L,t \to \infty\\ L/\Lth(t) = x}} \kb^{d=1}(t,L) - t = e^x - x - 1 
\equiv \kappab^{d=1}(x) 
\;,
\end{equation}
as given in Eq.~\eqref{eq:kb_scaling_d1} of the main text.

\subsection{Higher dimensional case}
For $d > 1$, the calculation of the Potts partition function poses a non-trivial problem. For $d = 2$, integrability can be used. However, here we are interested in the scaling limit where both the number of sites $\NN$ and the time $t$ are large. Let's set $z = e^{-\epsilon t}$ and consider the small $z$ (large $\epsilon$) expansion. We focus on $d=2$ for simplicity, but the procedure can be easily extended to any $d>1$.
This corresponds to a low-temperature expansion in the ferromagnetic phase. At the zero-th order, $\ZPotts$ is simply given by the $t$ possible groundstates. The leading correction is obtained changing the color of one site. This can be done on any site, choosing any among the $t-1$ remaining colors and will break $4$ bonds; the corresponding contribution to the partition function is therefore $\NN (t-1) z^4$. Higher orders in $z$ are obtaining changing colors at more sites. For instance, considering two site changes, we have two cases, according to whether the two sites are nearest neighbours or not. The first case gives a contribution $4 (t-1) \NN z^6 + 4 (t-1)(t-2) \NN z^7$. The second case gives instead $\NN(\NN-5)(t-1)^2 z^8$. In the limit, $L, t \to\infty$ with fixed $x = \NN t z^4 = \NN / \NN_{\rm Th}(t)$, we clearly see that only the last contribution survives. This corresponds to a dilute limit, which once accounting for the permutations among defects, takes in any $d>1$ the form 
\begin{equation}\label{eq:k_case_b_high_d}
    \lim_{\substack{L, t \to\infty \\ \NN/\NN_{\rm Th}(t) = x}} t^{-1} \kb^{d>1}(t,L) = \sum_{n=0}^\infty \frac{x^n}{n!} = e^x \equiv \kappab^{d>1}(x)  \;,
\end{equation}
with $x= \NN/ \NN_{\rm Th}$ $\NN_{\rm Th}(t) = z^{-2d} / t = e^{2 d \epsilon t}/t$, as given in Eq.~\eqref{eq:kb_scaling_dgt1} of the main text.


\section{SFF for translational invariant Floquet RPM}
In the TI Floquet RPM, the diagrams appearing in the expansion of the SFF can be put in correspondence of i) spatial permutations $\sigma$; ii) time translations at each site. We can thus face the problem in two steps, first fixing the permutation $\sigma$ and then summing over the possible choices of time translations at each site with fixed $\sigma$. We analyse separately the 1d from the $d>1$ case.

\subsection{One dimensional case}
As explained for the temporal random case, fixing the permutation $\sigma$ is equivalent to partitioning the $L$ sites into intervals $I_1, \ldots, I_n$ and then mapping the intervals on the top layer onto intervals on the bottom layer using cyclic permutations $\tilde \sigma \in M_n$ in \eqref{eq:Mndef}. For a fixed choice of the interval and of $\tilde \sigma$, the sum over the possible choices of the time translations can be still written in terms of a Potts-like partition function with modified Boltzmann weights: bonds inside the same interval are given by the matrix $\mathcal{W}$; bonds at the interface between two different intervals always give the trivial Boltzmann weight $e^{-\epsilon t}$. Since choosing a partition $I_1, \ldots, I_n$ can be done choosing the bonds where the edges of the intervals are, we can rewrite the SFF introducing the generalised partition function
\begin{equation}
    Z(\omega) =     \Tr[ (\mathcal{W} + \omega \mathcal{R})^L ] = \sum_{n=0}^{\infty} \omega^n Z_n \;,
\end{equation}
where we introduced the rank 1 matrix $\mathcal{R}_{\ttau, \ttau'} = e^{-\epsilon t}$. The coefficients $Z_n$ in the power series expansion of $Z(\omega)$ contain all the configurations where $n$ intervals are present and thus the trivial Boltzmann weight $\mathcal{R}_{\ttau,\ttau'}$ is used. We thus have
\begin{equation}\label{eq:kd_fin}
    K_{\rm TIF}^{d=1}(t, L) = L \sum_{n=0}^{L} a_n Z_n  \;,
\end{equation}
where, as before, $a_n$ is defined as the cardinality of the set in \eqref{eq:Mndef}.
The coefficients $Z_n$ can be computed explicitly from the spectrum of $\mathcal{W} + \omega \mathcal{R}$. Similarly to \eqref{eq:KFd1lambda}, we have
\begin{equation}
    Z(\omega) = (t-1)\lambda_-(t ,\omega)^L + \lambda_+(t, \omega)^L
\end{equation}
where $\lambda_-(t, \omega) = \lambda_-(t) = 1 - e^{- \epsilon t}$ and $\lambda_+(t, \omega) = 1 + e^{- t \epsilon} ((t-1) + \omega t)$. It follows that
\begin{equation}
    Z_0 = \kb^{d=1}(t,L)   \;, \qquad Z_n = \lambda_+(t, \omega=0)^L \binom{L}{n} \left(\frac{t}{e^{t \epsilon} + t -1}\right)^n \;, \quad n > 1 \;.
\end{equation}
Taking once again the scaling limit, we obtain
\begin{equation}
    \lim_{\substack{L, t \to \infty \\ L/\Lth(t) = x}} L^{-1} K_{\rm TIF}^{d=1}(t,L) - t = \kappab^{d=1}(x) + e^{x}(\kappac^{d=1}(x) - 1) = 
    - x - \ln (1 - x) \equiv
    \kappad^{d=1}(x)
\end{equation}
with  $x = L/\Lth(t)$ and $\Lth(t) = e^{t \epsilon}/t$,
as given in Eq.~\eqref{eq:kappaTIFd1} of the main text.
\subsection{Higher dimensional case}
An explicit formula for the SFF at finite $t$ and $L$ is hard to derive in this case, because it would require the computation of the Potts model partition function in $d$-dimensions where some bonds have been removed. We focus on the scaling limit $L,t \to \infty$ but with $x = \NN/\NN_{\rm Th}(t)$ kept constant. As explained above this correspond to a dilute limit. We thus have the expansion
\begin{equation}
    K_{\rm TIF}^{d>1}(t, L) \sim t \NN 
    \left[
    \sum_{n=0}^\infty \frac{(t\NN)^n}{n!}d_n e^{-2d \epsilon  n t} 
    \right]
    \left[ \sum_{m=0}^\infty \frac{\NN^m}{m!} (t-1)^n
    e^{-2 d  \epsilon m t}
    \right]
    \;.
\end{equation}
The origins of the expression are explained as follows: In the first square bracket, the sum over $n$  accounts for the deranged defects on top of the rigid spatial translation which can be chosen in $\binom{N}{n} \sim N^n / n!$ ways. The factor of $t$ accounts for $t$ choices of the $\ttau$ variables for these defects, and the factor $d_n$ counts their derangements. 
In the second square bracket, the sum over $m$ accounts for the defects in the sum over the $\ttau$ variables. We can choose $m$ defects in $\binom{\NN}{m} \sim \NN^m / m!$ ways and each of them can be independently changed with the $t-1$ other colours different from the background. Taking again the scaling limit 
\begin{equation}
    \lim_{\substack{L, t \to \infty \\ \NN/\NN_{\rm th}(t) = x}} (t \NN)^{-1} K_{\rm TIF}^{d>1}(t,L) = \kappab^{d>1}(x) \times \kappac^{d>1}(x) = \frac{1}{1-x} \equiv \kappad^{d>1}(x) \;,
    \end{equation}
with $\NN_{\rm Th}(t) = e^{2 d \epsilon t}/t $ and $x = \NN /\NN_{\rm Th}(t)$, 
as given in Eq.~\eqref{eq:kappaTIFdgt1} of the main text.


\section{Generalised unit cell in large-$q$}
\begin{figure}[ht]
	\includegraphics[width=0.4\columnwidth
	]{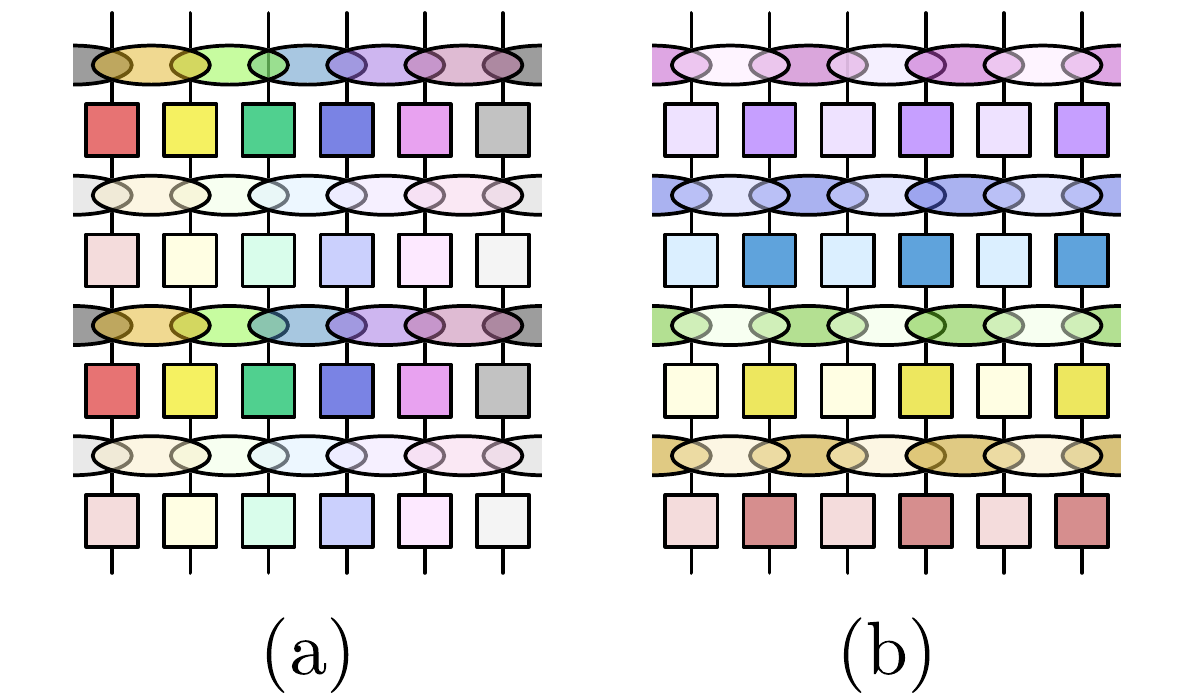}
	\caption{
Illustrations of 
	(a) Floquet (and spatially random) RPM which is invariant under 2-discrete-time translation;
and	(b)  TI (and temporally random) RPM which is invariant under 2-site translation.
For each case, gates of the same colors are identical.
	}\label{fig:large_unit_cell}
\end{figure}

In this appendix, we demonstrate the robustness of the scaling forms. To this end, we compute the scaling forms in the large-$q$ limit of the modified circuit models, namely TI RPM (Floquet RPM) that are invariant under $p$-site ($p$-discrete-time) translation in space (time). The 2-site and 2-discrete-time translational invariant model is illustrated in Fig. \ref{fig:large_unit_cell}a and b respectively. 
A summary of the results is as follows. For TI RPM with $p$-site translational invariance, we show that for $d=1$, the scaling form $\kappac^{(p)}$ is lower bounded by $\kappac^{(1)}$ in Eq. \eqref{eq:KTIRPM_psite_bound}. 
For $d \geq 2$, we show that $  \kappac^{(p)} =  \left[ \kappac^{(1)}(x) \right]^p $ in Eq. \eqref{eq:kc_higherd_2site}. 
For Floquet RPM with $p$-discrete-time translational invariance, we show that for all dimension $d$ that 
$ \kappab^{(p)}(x) = \kappab^{(1)}(x) $ in Eq. \eqref{eq:kb_scaling_ptime} and \eqref{eq:kb_scaling_ptime_higherd}.

\subsection{SFF for TI RPM with $p$-site translational invariance}
\subsubsection{One-dimensional systems}
For $d=1$, we derive a lower bound for $\kc^{(p)}$ (and $\kappac^{(p)}$) for TI RPM with $p$-site translational invariance. 
We will first present the derivation for $p=2$ then generalise it to any $p \in \mathbb{N}^+$.
Following the derivation of Eq. \eqref{eq:sffmulti}, we have for $p=2$, a sum over two permutations, $\sigma_1$ and $\sigma_2$
\begin{equation}
\label{eq:KTIRPM_2site}
\kc^{(2)} (t, \Leff)
=
\sum_{\substack{
\sigma_1 \in \mathcal{S}_{\rm odd} \\ 
\sigma_2 \in \mathcal{S}_{\rm even} 
}} e^{-\epsilon t 
 \left( L -  \Npb{ (\tilde{\sigma}[\sigma_1, \sigma_2] )}  \right)}  \;,
\end{equation}
where $\Leff = L/2$ is the number of unit cells, $\mathcal{S}_{\rm odd}$ / $\mathcal{S}_{\rm even}$  
are permutations restricted to the odd/even sublattices and 
$\Npb(\sigma)$ is the number of preserved bonds as defined in the main text. The permutation $\tilde{\sigma}$ is defined composing $\sigma_1$ and $\sigma_2$  
\begin{align}
\label{eq:tilde_sigma}
\tilde{\sigma}[\sigma_1, \sigma_2] (i) = 
\begin{cases}
\sigma_1 (i)   & \text{if }i\text{ is odd}\\
\sigma_2 (i)   &  \text{if }i\text{ is even}
\end{cases}
\end{align}
It is difficult to evaluate \eqref{eq:KTIRPM_2site} exactly, but a lower bound can be identified. Observe that one can always restrict the broken bonds to be placed in between site $i$ and $i+1$ for even $i$. This means $\kc^{(2)}> \Leff \sum_{n=0}^{\Leff} A_1 (\Leff, n)$, where $A_1$ is given in Eq. \eqref{eq:case_c_1d_identity} in the main text. In addition to the above configurations, one can find another set of diagrams where the broken bonds are placed between site $i$ and $i+1$ with odd $i$.  
Putting the two cases together and accounting for the overcounting of the cases where no bonds are broken we have
\begin{align}
\label{eq:KTIRPM_2site_bound}
\kc^{(2)}(t,\Leff ) / \Leff
& \geq 
\sum_{n_1 =0}^{\Leff} \binom{\Leff}{n_1} a_{n_1} e^{-n_1 \epsilon t} 
+
\sum_{n_2 =0}^{\Leff} \binom{\Leff}{n_2} a_{n_2} e^{-n_2 \epsilon t} 
-1
\end{align}
This expression is easily generalised to arbitrary $p$ to 
\begin{align}
\label{eq:KTIRPM_psite_bound0}
\kc^{(p)}(t,\Leff ) / \Leff
 \geq 
p \sum_{n =0}^{\Leff} \binom{\Leff}{n} a_{n} e^{-n \epsilon t} 
-(p-1)
\end{align}
with $\Leff = L/p$ in general.

In the scaling limit where $t,L \to \infty$ with fixed $x = \Leff / \Lth = \Leff e^{-\epsilon t}$, we define
\begin{equation}
\kappac^{(p)}(x) \equiv \lim_{\substack{L,t \to \infty\\ \Leff/\Lth(t) = x}} \Leff^{-1} \kc^{(p)}(t,L) \; .
\end{equation}
Note that with this definition $\kappac^{(p)}(x = 0) = 1$, which corresponds to the RMT behavior at large times. From \eqref{eq:KTIRPM_psite_bound0}, we obtain the lower bound for the scaling function
\begin{equation}
\label{eq:KTIRPM_psite_bound}
\kappac^{(p)}(x) 
\geq 
p  \kappac^{(1)} (x) -(p-1)
\;, \qquad 
\end{equation}
Note that this equation implies the existence of a divergence at $x=1$ for all $p$'s.
\subsubsection{Higher-dimensional systems}
For $d>1$, with the foresight that we will use the scaling limit where only dilute deranged defects matter, we define a generic higher-dimensional TI RPM as follows:
Consider a TI RPM made up of super-sites of $p$ qudits, each of which evolve under independently drawn CUE, and couple with $z$ other qudits under the (independently drawn) random phase gates defined in the main text. 
Additionally, we demand that the translational invariance of the supersites is preserved.
Consider the scaling limit where $t,L \to \infty$ at fixed $x= \NNeff / \NN_{\mathrm{Th}}$ with $\NNeff=\NN / p $ and  $\NN_{\mathrm{Th}} = e^{z\epsilon t}$.
In this limit, SFF is a sum over deranged defect diagrams labelled by $(k_1, k_2, \dots, k_p)$ where $k_i$ is the number of deranged dilute defects of the $i$-th type of qudits in the supersite. 
Following the derivation of Eq. \eqref{eq:sffmulti}, \eqref{eq:AlargeLdgt1} and \eqref{eq:kc_scaling_dgt1}, in the limit of large $\NN$, we write
\be
\NNeff^{-1} \kc^{(p)} \sim \prod_{i=1}^p  \sum_{k_i=0}^{\NNeff} \frac{ d_{k_i}  \NNeff^{k_i}}{k_i! \, \NNTh^{k_i}} = \left[ \NNeff^{-1} \kc^{(1)} \right]^p \;,
\ee
where again $d_k$ is the number of derangements of $k$ elements. 
Using the definitions 
\begin{align} 
    \kappac^{(p)}(x) & \equiv \lim_{\substack{\NN,t \to \infty\\ \NNeff /\NN_{\rm Th}(t) = x}} \NNeff^{-1} \kc^{(p)}(t,L) \; ,
\end{align}
we arrive a more compact statement in the scaling limit
\begin{equation} \label{eq:kc_higherd_2site}
    \kappac^{(p)} =  \left[ \kappac^{(1)}(x) \right]^p  \;. 
\end{equation}
Again, with this convention we have  $\kappac^{(p)}(x = 0) = 1$,  which corresponds to the RMT behavior at large times.

\subsection{SFF for Floquet RPM with $p$-discrete-time translational invariance}
The calculation of SFF for RPM with $p$-discrete-time translation invariance closely follows the one for RPM with 1-discrete-time translation invariance (see main text and \cite{cdc1, cdc2}).
For all dimensions $d$, upon averaging over the CUE in the large-$q$ limit following \cite{cdc1, cdc2}, SFF is mapped to a Potts model with DOF $\ttau_i$ at each site $i$.  $\ttau_i$ can take $\teff = t/p$ number of possible states, corresponding to the $\teff$ possible ladder diagrams. 
The averages over the random phases give an effective  Boltzmann weight $\mathcal{W}_{\ttau, \ttau'}  = e^{- \teff(1-\delta_{\ttau, \ttau'}) \epsiloneff}$ with $\epsiloneff = \epsilon /p$ for nearest neighbour pairs of state $\ttau$ and $\ttau'$.
Therefore, the SFF $\kb^{(p)}(\teff, L) $ can be written as the partition function of a $\teff$-state Potts model with Boltzmann weight parametrized by $\epsiloneff$.  

\subsubsection{One-dimensional systems }
In $d=1$, we define in the scaling limit,
\begin{align} 
    \kappab^{(p)}(x) & \equiv \lim_{\substack{L, t \to\infty\\ \NN/\NN_{\rm Th}(t) = x}} \kb^{(p)}(\teff , L) - \teff
    \;.
\end{align}
where $L_{\mathrm{Th}}(\teff) =   e^{\teff \epsiloneff}/ \teff = p e^{ t \epsilon}/ t$.  $\kappab^{(p)}(x) $ can now be related to $\kappab^{(1)}(x) $ as
\begin{equation}
\label{eq:kb_scaling_ptime}
\kappab^{(p)}(x)
= e^x - x - 1 = \kappab^{(1)}(x)\;.
\end{equation}
In other words, after properly defining the Thouless length, the scaling function is the same independently of $p$. Note that at large time, we have  $\kappab(x = 0) = 0$, and the RMT result is reproduced.

\subsubsection{Higher-dimensional systems }
For $d\geq 1$, we define $\NNTh = e^{z \epsiloneff \teff} / \teff$ with $\teff = t/p$ and $\epsiloneff = p\epsilon$. Furthermore, we define
\begin{align} 
    \kappab^{(p)}(x) & \equiv \lim_{\substack{L, t \to\infty\\ \NN/\NN_{\rm Th}(t) = x}} \teff^{-1} \kb^{(p)}(\teff , L)
    \;.
\end{align}
In the scaling limit,  $\kappab^{(p)}$ coincides with $\kappab^{(1)}$ as 
\begin{equation}
\label{eq:kb_scaling_ptime_higherd}
 \kappab^{(p)}(x) = \kappab^{(1)}(x) \;,
\end{equation}
i.e. the scaling function is independent of $p$. Note again that at large time, we have  $\kappab(x = 0) = 0$, and the RMT result is reproduced.


\section{Brick wall model (BWM) \label{app:BWM}}
The one-dimensional BWM is defined by a quantum circuit which is a matrix product 
\be
W(t) = \prod_{t'=1}^t w(t') 
\ee
where  $w(t)= w_2(t)\,  w_1(t) $ is a $q^L\times q^L$ operator. 
\begin{equation}
\label{eq:W1_bwm}
w_1(t) = \bigotimes_{n=1}^{L/2}
u_{2n-1,2n}(t)
\end{equation}
is a tensor product of $q^2 \times q^2$ unitary matrices $u_{2n-1,2n}(t)$ chosen from the circular unitary ensemble (CUE) and acting on site $2n-1$ and $2n$. 
\begin{equation}
\label{eq:W2_bwm}
w_2(t) = \bigotimes_{n=1}^{L/2}
u_{2n,2n+1}(t)
\end{equation}
is again a tensor product of $u_{2n,2n+1}(t)$ drawn from CUE except that the unitary gate acts on site $2n$ and $2n+1$. The model is defined with periodic boundary condition with $u_{L,L+1}(t)$ acting on site $L$ and $1$.

For the temporally and spatially random BWM, each unitary gate $u_{n,n+1}(t)$ is drawn independently. 
For Floquet BWM, we take unitary gates acting on different pairs of sites to be independently drawn, while gates acting on the same pair of sites at different discrete time $t$ to be identically drawn, i.e. $w(t) = w(t')$ for $t\neq t'$ and $W(t) = w^t$. 
For TI BWM, gates in $w_1(t)$ (and separately in $w_2(t)$) acting on different pairs of sites at the same time are identically drawn, but gates acting on different discrete times are independently drawn. 
For Floquet TI BWM, $u_{n,n+1}(t) = u_{n',n'+1}(t')$ for even integer $n$ and for all $t$. The equation separately holds true for odd integer $n$.



\section{Numerical methods and results}
We simulate the RPM with $q=3$, $\epsilon=2$ and $d=1$, and the BWM with $q=2,3$, both in $d=1$ and with periodic boundary conditions. 
(We find that RPM with $q=2$ and large $\epsilon$ is not fully chaotic in the sense that the linear ramp does not appear in late $t$.)
We consider two types of numerical simulations: ``Time-direction'' simulations 
and ``space- (or dual-)direction'' simulations, which involve performing calculations by acting matrices in the Hilbert space $\mathcal{H} = \bigotimes^L \mathbb{C}^q$ and the dual Hilbert space $\tilde{\mathcal{H}} = \bigotimes^t \mathbb{C}^q$  respectively. Although Figure \ref{fig:collapse} uses only dual direction numerics, here we discuss both the directions.

For the time-direction simulations with translational invariant (but not Floquet) circuits, the computation of $K(t,L)$ vs $t$ for different $L$ involves sparse matrix multiplication of locally-supported random unitaries, and taking the trace. 
For translational invariant Floquet circuits, we compute $K(t,L)$ for a fixed $L$ with exact diagonalizations (ED).

For the space-direction simulations, we define a dual Floquet operator $V$, which
we explicitly construct for RPM. $V$ for BWM can be similar constructed following the procedure below.
Informally, $V$ is taken to be the tensor product of the first column of tensors in Fig. \ref{fig:model_rpm}b and \ref{fig:model_rpm}c. 
More precisely, we introduce the computational basis $\bv = \{b^1, \ldots, b^t\}$ with each $b^\mu= 1,\ldots,q$. The dual tensor for RPM can now be written as 
\begin{subequations}
\label{eq:VVsdef}
\begin{align}
&[\VV_{1}]_{\bv,\bv'} = \prod_{\mu=1}^t e^{\imath \varphi_{b^\mu, {b^\mu}'} } \label{eq:VV1}  \;, \\
&[\VV_{2}]_{\bv,\bv'} =  \prod_{\mu=1}^t u_{b^{\mu+1}, b^{\mu}} \delta_{\bv,\bv'} 
\;,
\end{align}
\end{subequations}
so that $V = \VV_{2} \VV_{1}$ and $V(L) = V^L$.
Note that in the dual formulation the $1$-body unitary matrices in $\TT_1$ are converted into $2$-body diagonal matrices in $\VV_2$, while the $2$-body phases in $\TT_2$ are converted into the $1$-body $\VV_1$.
Because of periodic boundary conditions and taking the trace, the SFF for the dual Floquet operator can also be expressed as a trace,
\begin{equation}
K(t,L)=\average{\Tr_{\mathcal{H}}[W(t)] \Tr_{\mathcal{H}}[W^\dagger(t)]}=\average{\Tr_{\tilde{\mathcal{H}}}[V(L)] \Tr_{\tilde{\mathcal{H}}}[V^\dagger(L)]} \;, 
\end{equation}
where we have added subscripts to the traces to emphasize the Hilbert spaces on which matrices $W$ and $V$ act. The dual tensors for TI BWM, with a computational basis $\bv = \{b^1,b^2 \ldots, b^{2t}\}$, can be similarly written as 

\begin{subequations}
\label{eq:VVsdef_bwm}
\begin{align}
&[\VV_{1}]_{\bv,\bv'} = \prod_{\mu=1}^t \tilde{u}_{b^{2\mu}, b^{2\mu+1}}(\mu) \;, \\
&[\VV_{2}]_{\bv,\bv'} =  \prod_{\mu=1}^t \tilde{u}_{b^{2\mu-1}, b^{2\mu}}(\mu) 
\;,
\end{align}
\end{subequations}
where $\tilde{u}_{ij,kl}=u_{jl,ik}$, $u_{ij,kl}$ being a CUE random matrix acting on two sites, and $\tilde{u}$ is the dual of the unitary gate, which is non-unitary in general. The dual Floquet operator is given by $V = \VV_{2} \VV_{1}$ and $V(L) = V^{L/2}$.

\subsection{SFF for translational invariant circuits
\label{app:sfftransl}}
\begin{figure}[ht]
\begin{minipage}[t]{0.31\textwidth}
\includegraphics[width=1.1\linewidth,keepaspectratio=true]{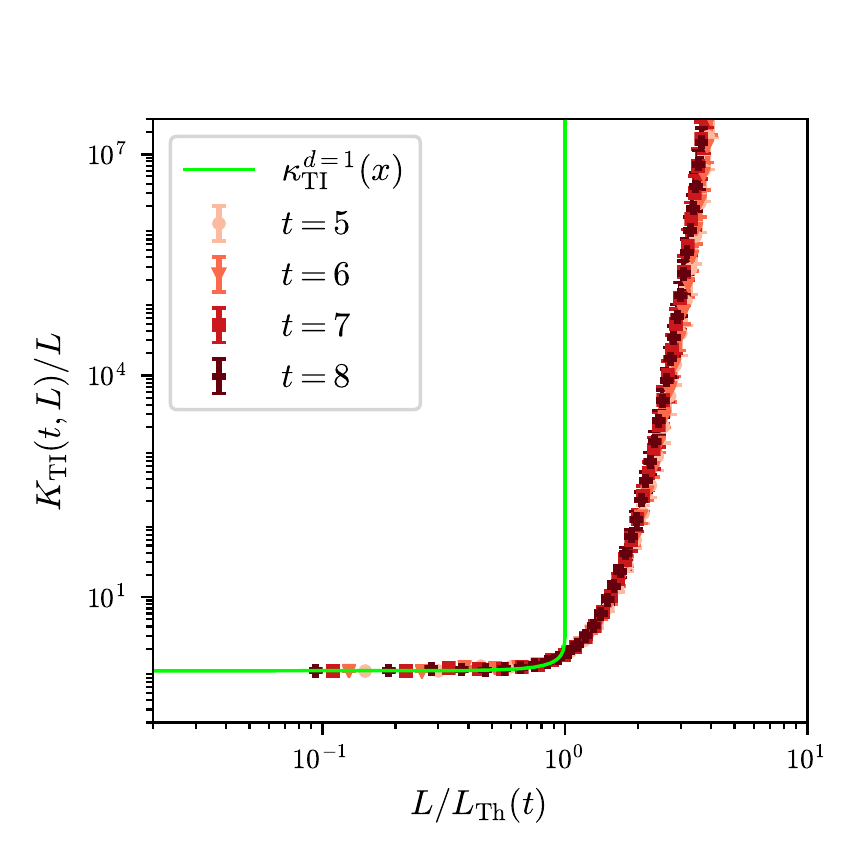}
\end{minipage}
\hspace*{\fill} 
\begin{minipage}[t]{0.31\textwidth}
\includegraphics[width=1.1\linewidth,keepaspectratio=true]{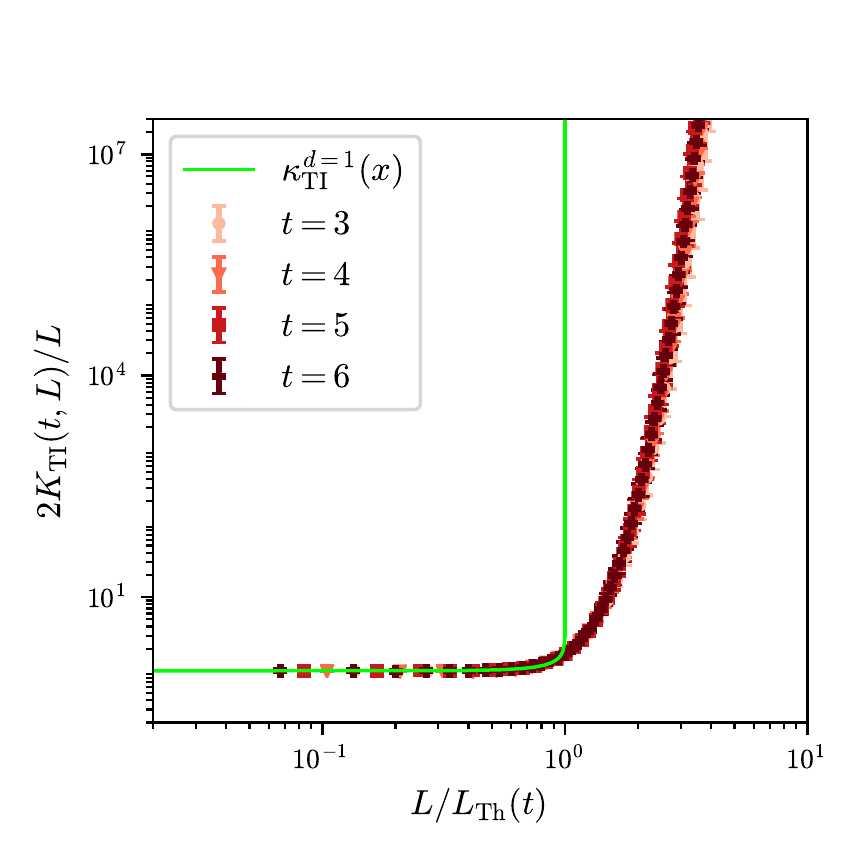}
\end{minipage}
\hspace*{\fill} 
\begin{minipage}[t]{0.31\textwidth}
\includegraphics[width=1.1\linewidth,keepaspectratio=true]{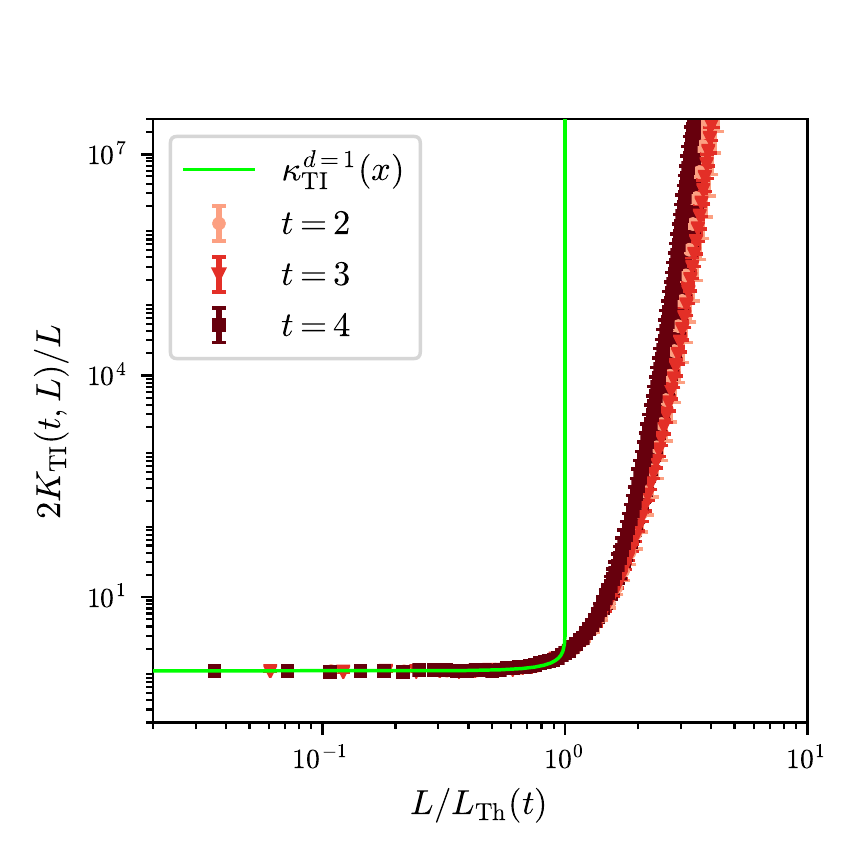}
\end{minipage}
\caption{
 $K(t,L)/L$ vs. $L/ \Lth$ for TI-RPM at $q=3$ (left), TI-BWM at $q=2$ (middle), and TI-BWM at $q=3$ (right).
    } \label{fig:caseC_collapse}
\end{figure}

We elaborate on the middle panel of Figure \ref{fig:collapse}, and provide further evidence of consistency with the predicted scaling form in Eq. \ref{eq:kc_scaling}.
For both RPM and BWM, $K_{\mathrm{TI}}(t,L)$ is computed by exact diagonalizing the dual Floquet operator $V$. In figure \ref{fig:caseC_collapse}, we re-plot the data for RPM (left), BWM with $q=2$ (middle), and also add the data for BWM with $q=3$ in the right panel (whose importance is spelled out in the discussion on TIF circuits), for which only three system sizes were accessible. Averaging was done over $12000-15000$ realizations of $V$ for RPM, $8000-15000$ realizations for BWM with $q=2$, and over $4000-10000$ realizations for BWM with $q=3$.

\begin{figure}[ht]
\begin{minipage}[t]{0.48\textwidth}
\includegraphics[width=0.8\linewidth,keepaspectratio=true]{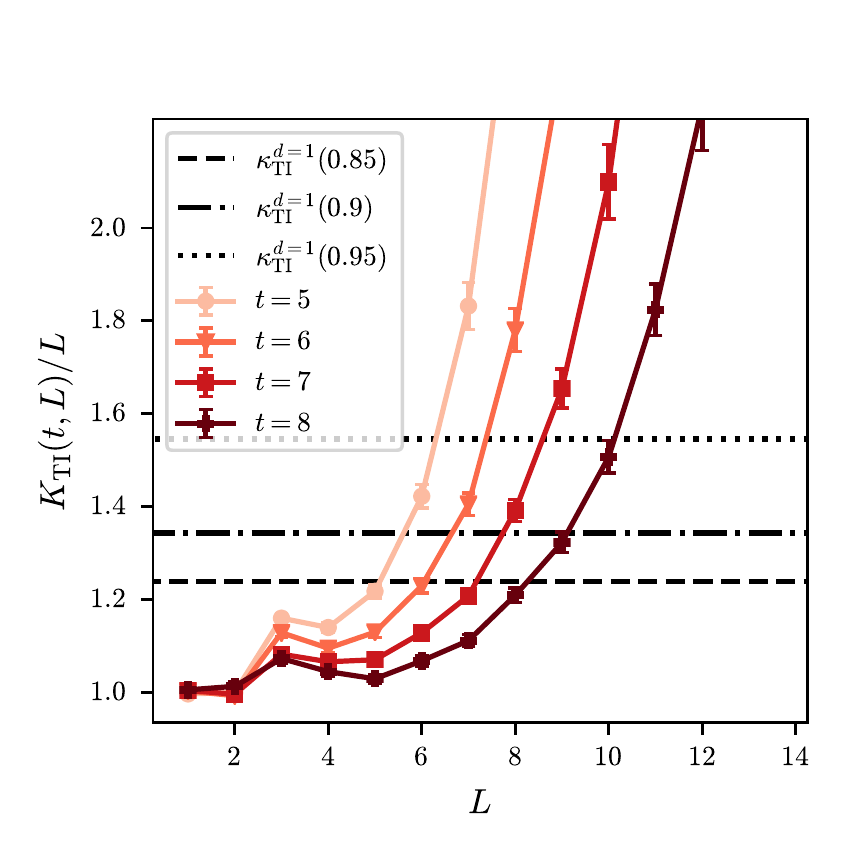}
\end{minipage}
\hspace*{\fill} 
\begin{minipage}[t]{0.48\textwidth}
\includegraphics[width=0.8\linewidth,keepaspectratio=true]{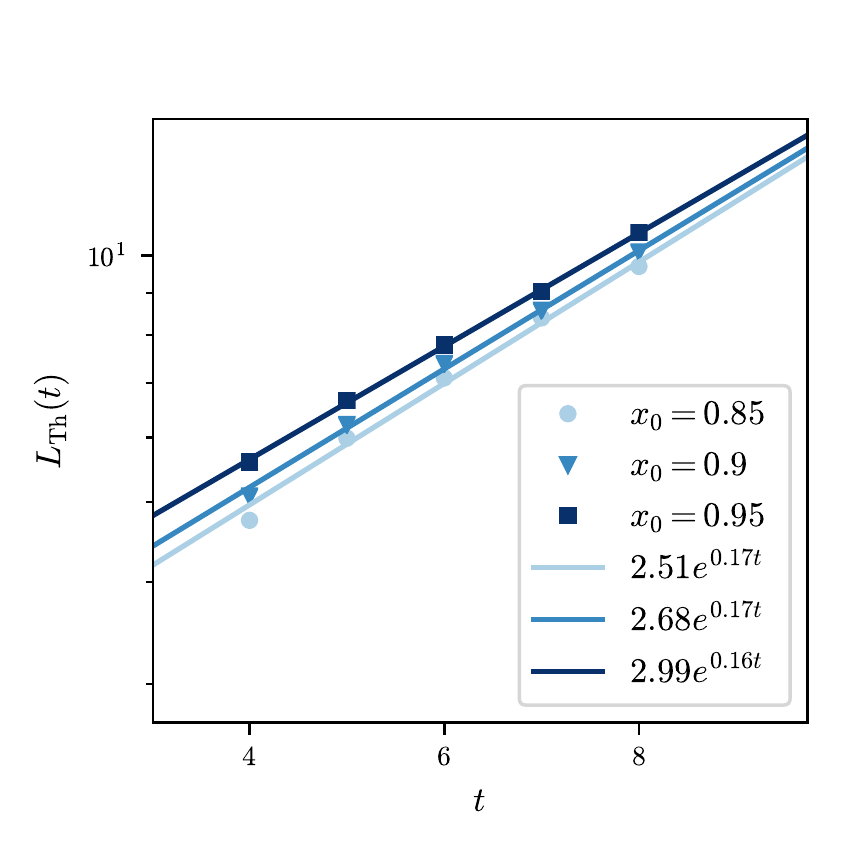}
\end{minipage}

\caption{
Left : $K_{\mathrm{TI}}(t,L)/L$ vs $L$ for different $L$  using space direction simulations of TI RPM, with the intersections $\tilde{L}(t)$ with constant horizontal lines $\kappac^{d=1}(x_0)$ for a few $x_0$ used to compute $\Lth(t)=\tilde{L}/x_0$  . Right : $\Lth(t)$ vs $t$ for different choices of $x_0$. 
	The corresponding exponential fits (solid lines) show reasonable agreement with the expected form from Eq. \ref{eq:kc_scaling} ($\Lth=e^{\epsiloneff t}$), with some effective $\epsiloneff \neq 2$ due to finite-$q$ effects.
    } \label{fig:L_th_rpm_caseC}
\end{figure}

Now we describe the details of how we obtain $\Lth (t)$ and plot the collapse in Figure \ref{fig:collapse}.
The  horizontal axes in \ref{fig:collapse} are all scaled such that the finite-$q$ numerics for $K_{\mathrm{TI}}(t,L)/L$ is equal to the infinite-$q$ scaling function $\kappac^{d=1}(x)$ (Eq. \ref{eq:kc_scaling}) at a specific point $x_0<1$. 
We choose $x_0=0.95$ and determine $\tilde{L}(t)$ for a given $t$ such that $K_{\mathrm{TI}}(t,\tilde{L})/\tilde{L}=\kappac^{d=1}(x_0)$. 
Since $x_0=\tilde{L}(t)/L_{\mathrm{Th}}(t)$ by definition, we get the corresponding Thouless time $L_{\mathrm{Th}}(t)=\tilde{L}(t)/x_0$. 
Lastly, we rescale the horizontal axis to be $x=L/L_{\mathrm{Th}}(t)$, so the numerics and the scaling function can be compared directly. 
We visually show this procedure in Figure \ref{fig:L_th_rpm_caseC}, where the left panel shows the horizontal lines for a few $x_0$ that we draw to extract $\tilde{L}(t)$, and the right panel shows  $\Lth(t)$ vs $t$ for those $x_0$. We find little difference between the behaviour at different $x_0$, and exponential fits to $\Lth(t)$ are consistent with the expected scaling form in Eq. \ref{eq:kc_scaling}.

$\Lth(t)$ for BWM (both with $q=2$ and $q=3$) were similarly calculated and is shown in the middle and the right panel of Figure \ref{fig:L_th_all}.

\begin{figure}[ht]
\begin{minipage}[t]{0.31\textwidth}
\includegraphics[width=1.1\linewidth,keepaspectratio=true]{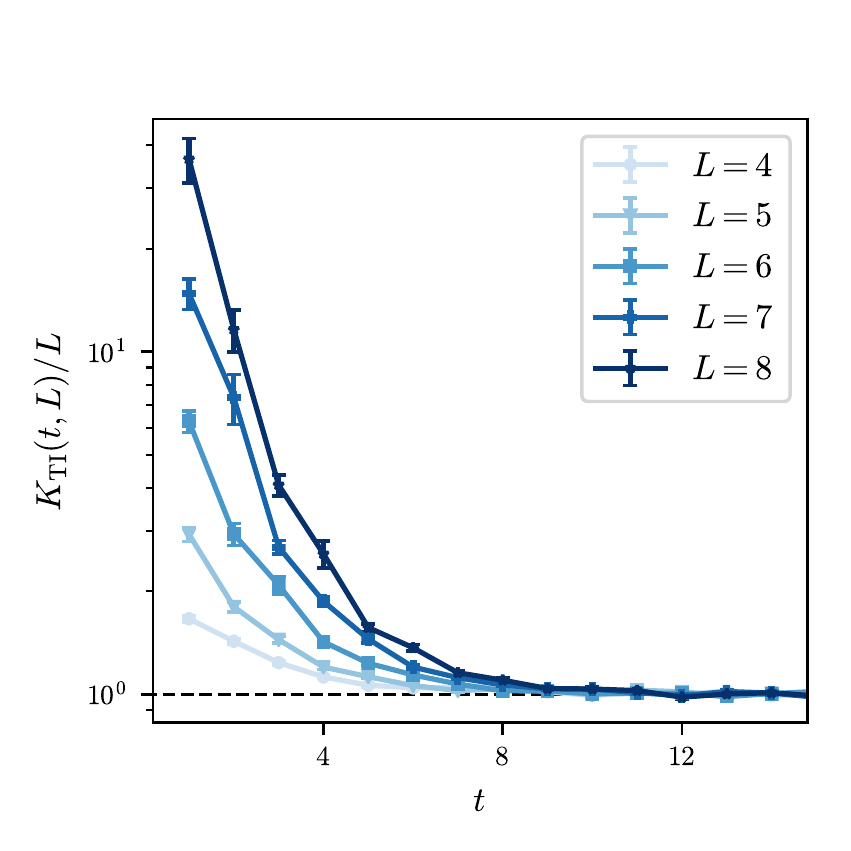}
\end{minipage}
\hspace*{\fill} 
\begin{minipage}[t]{0.31\textwidth}
\includegraphics[width=1.1\linewidth,keepaspectratio=true]{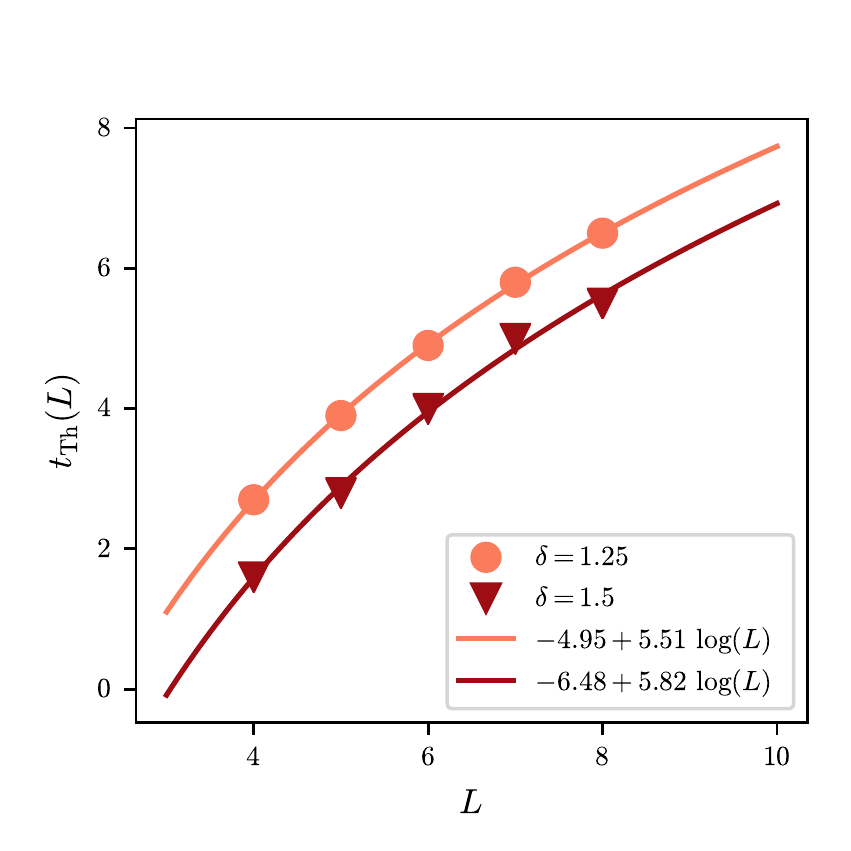}
\end{minipage}
\hspace*{\fill} 
\begin{minipage}[t]{0.31\textwidth}
\includegraphics[width=1.1\linewidth,keepaspectratio=true]{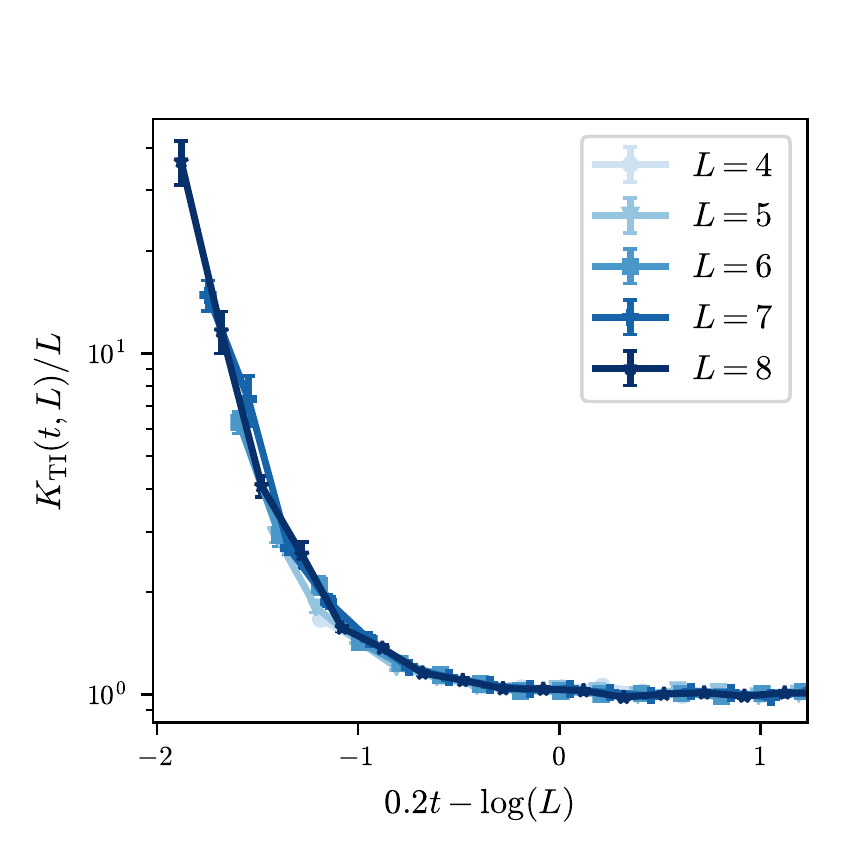}
\end{minipage}
\caption{
Left : $K_{\mathrm{TI}}(t,L)/L$ vs $t$ for different $L$ obtained using time direction simulations of TI RPM. Middle : $\tth(L)$ vs $L$, computed by numerically solving $K_{\mathrm{TI}}(\tth,L)/L =\delta$. A logarithmic fit is consistent with the expectation that $\Lth(t)=Le^{-\epsiloneff t}$, with an effective $\epsiloneff  \neq 2$ due to finite-$q$ corrections, and also consistent with the fit in Fig.  \ref{fig:L_th_rpm_caseC}. Right : $K_{\mathrm{TI}}(t,L)/L$ vs $0.2t-\log(L)$.
    } \label{fig:t_dir_rpm}
\end{figure}

As additional checks for consistency, we consider time direction simulations, using sparse matrix multiplication and trace evaluation, in Figure \ref{fig:t_dir_rpm}, where the left panel shows $K_{\mathrm{TI}}(t,L)/L$ vs $t$ for different $L$, averaged over $6000-9000$ realizations of $W(t)$. Using this data, $t_{\mathrm{Th}} (L)$ can be obtained analogous to $L_{\mathrm{Th}}(t)$ by looking at the intersection of $K_{\mathrm{TI}}(t,L)/L$ with a constant $\delta=\kappac^{d=1}(x_0)$ for some $x_0$, as plotted in the middle panel. Note that here we are only interested in the scaling form of $t_{\mathrm{Th}}(L)$ hence we ignore the normalization arising due to different $\delta$. Good logarithmic fits indicate consistency with the form of $\Lth(t)=Le^{-\epsiloneff t}$, where the fitted $\epsiloneff \neq 2$ is due to finite $q$ effects. The scaling form is then further corroborated in the right panel, showing a collapse with $0.2t-\log(L)$, with $\epsiloneff$ being consistent across all plots.

\subsection{SFF for Floquet circuits}
\begin{figure}[ht]
\begin{minipage}[t]{0.31\textwidth}
\includegraphics[width=1.1\linewidth,keepaspectratio=true]{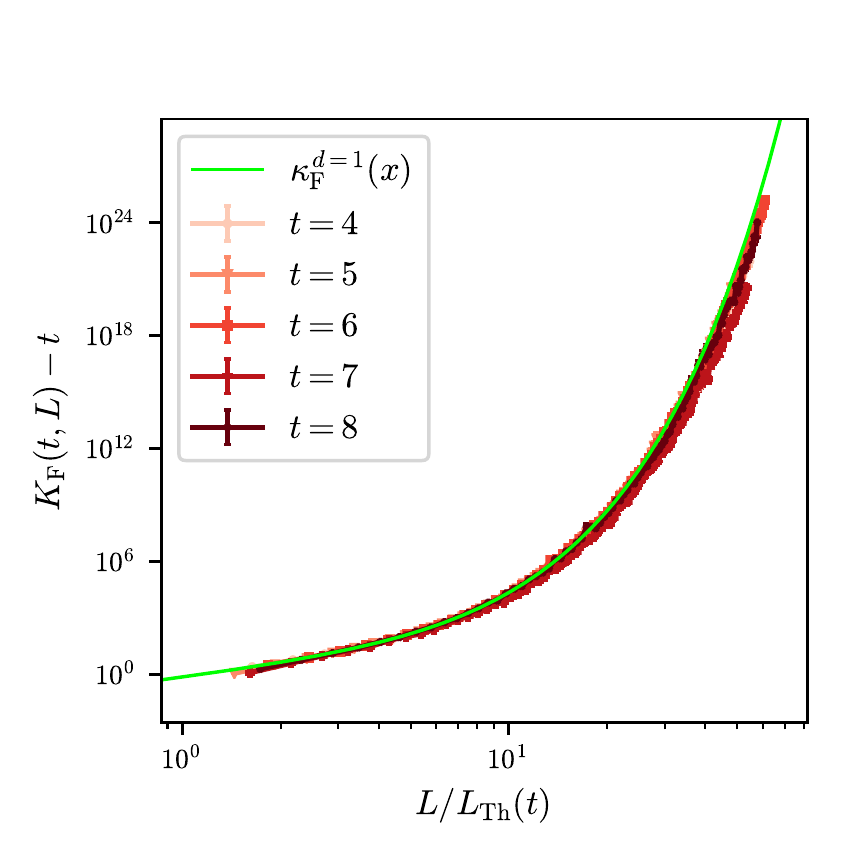}
\end{minipage}
\hspace*{\fill} 
\begin{minipage}[t]{0.31\textwidth}
\includegraphics[width=1.1\linewidth,keepaspectratio=true]{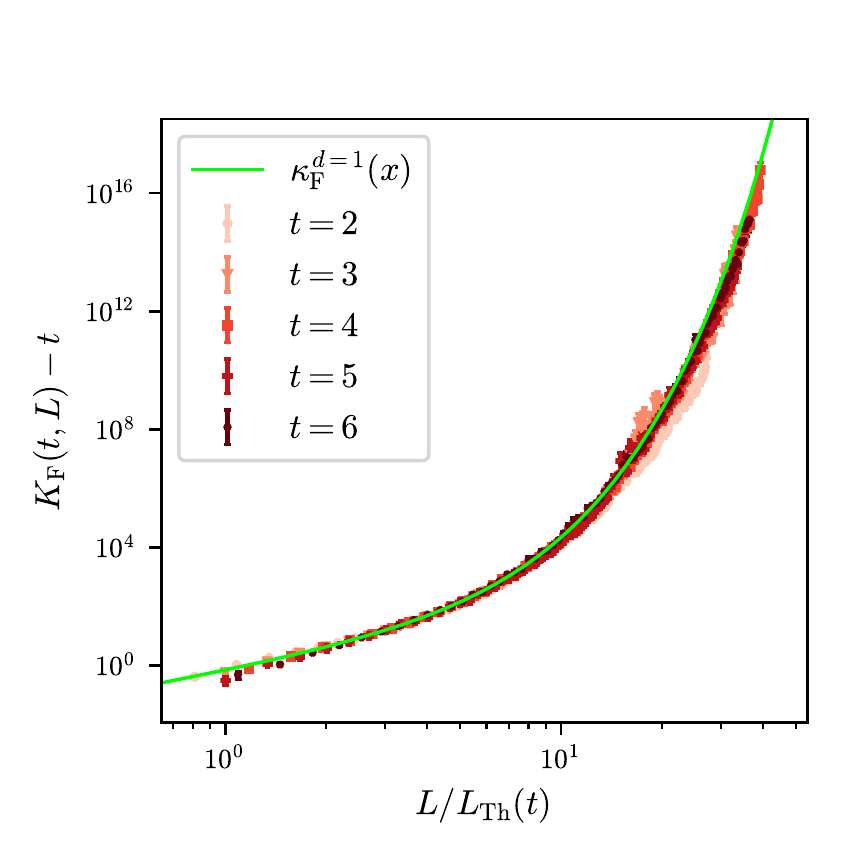}
\end{minipage}
\hspace*{\fill} 
\begin{minipage}[t]{0.31\textwidth}
\includegraphics[width=1.1\linewidth,keepaspectratio=true]{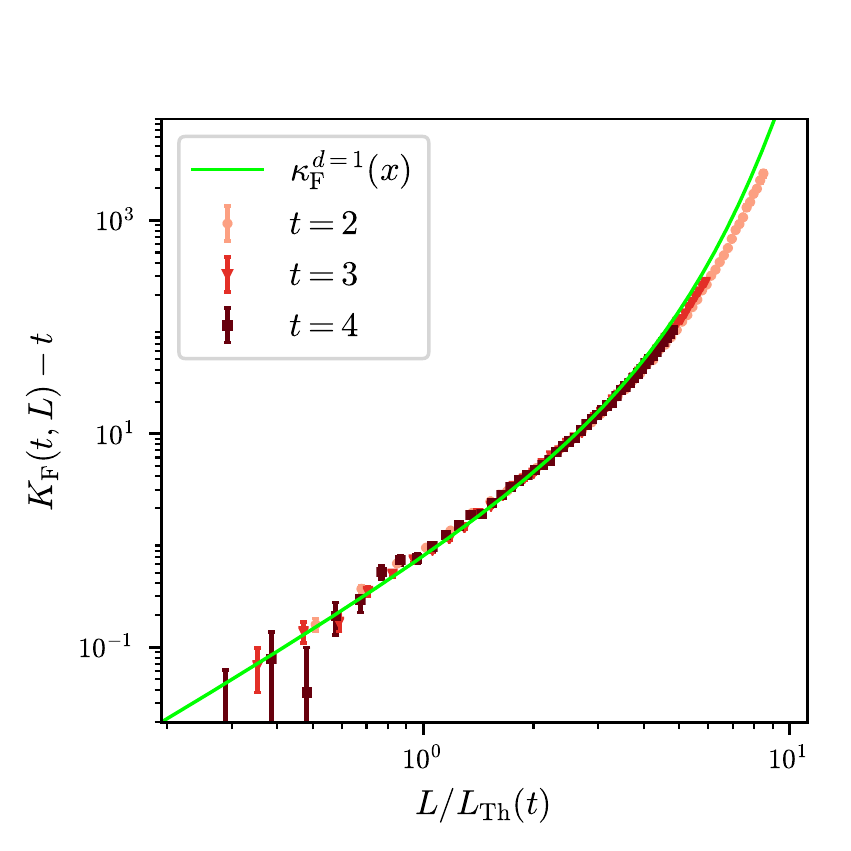}
\end{minipage}
\caption{
 $K(t,L)-t$ vs. $L/ \Lth$ for F-RPM at $q=3$ (left), F-BWM at $q=2$ (middle), and F-BWM at $q=3$ (right). 
    } \label{fig:caseB_collapse}
\end{figure}

Here, we elaborate on the left panel of Figure \ref{fig:collapse}. We first show in Figure \ref{fig:caseB_collapse} the collapse of $K_{\mathrm{F}}(t,L)-t$ vs $L/\Lth(t)$ separately for RPM with $q=3, \epsilon=2$ (left), BWM with $q=2$ (middle), and additionally BWM with $q=3$(right); and we find that for all these circuits, the infinite-$q$ scaling function in Eq. \ref{eq:kb_scaling_d1} is in excellent agreement with finite-$q$ numerics for these circuits. 
Note that these are all space direction simulations performed through sparse matrix multiplication and trace computation and not through ED. The data for RPM has been averaged over $4000-10000$ realizations, and for BWM with $q=2$ over $3000-10000$ realizations, while for BWM with $q=3$, we averaged over $4000-10000$ realizations. 
$\Lth(t)$ were obtained in a similar fashion as for TI RPM (Figure \ref{fig:L_th_rpm_caseC}), fixing $x_0=3$, and are shown in the left panel of Figure \ref{fig:L_th_all}.

\subsection{SFF for translational invariant Floquet circuits}
\begin{figure}[ht]
\begin{minipage}[t]{0.31\textwidth}
\includegraphics[width=1.1\linewidth,keepaspectratio=true]{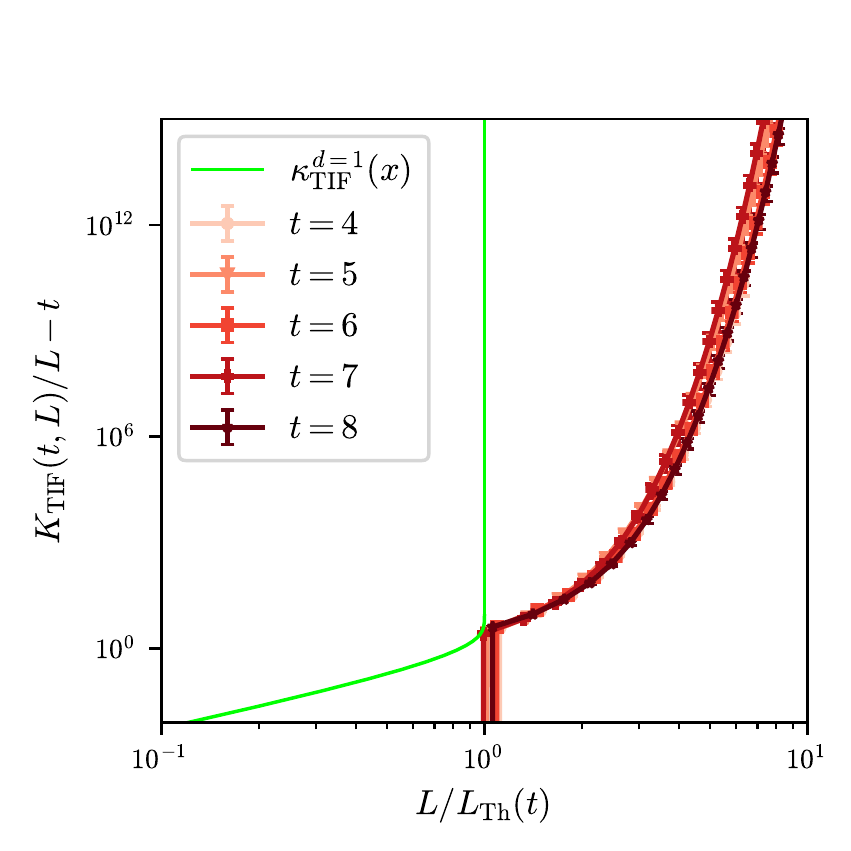}
\end{minipage}
\hspace*{\fill} 
\begin{minipage}[t]{0.31\textwidth}
\includegraphics[width=1.1\linewidth,keepaspectratio=true]{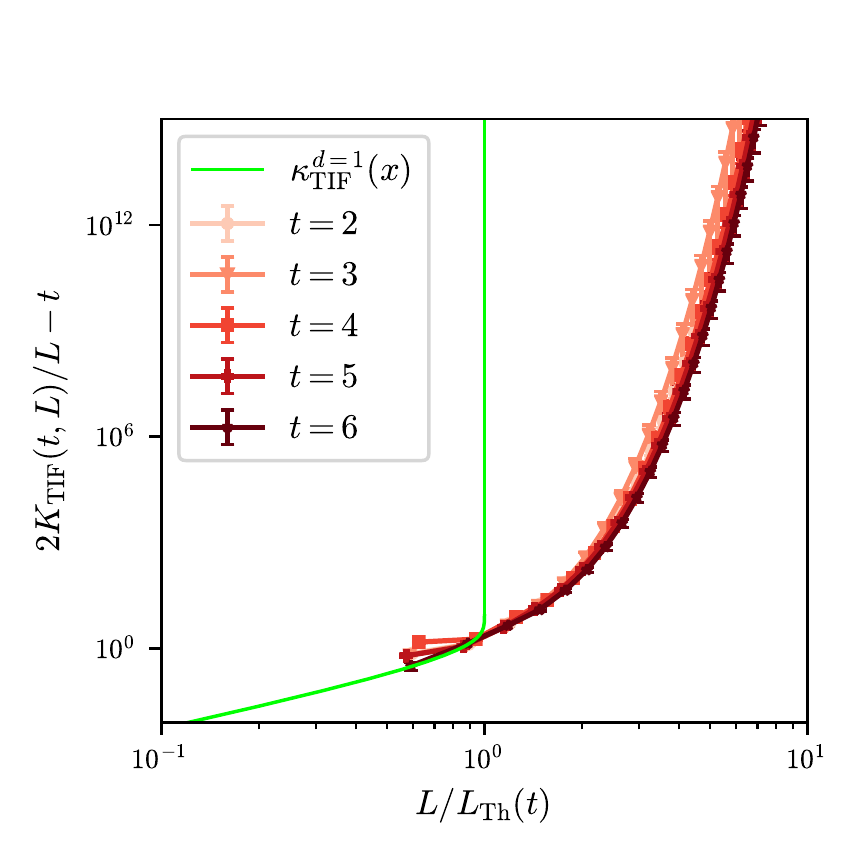}
\end{minipage}
\hspace*{\fill} 
\begin{minipage}[t]{0.31\textwidth}
\includegraphics[width=1.1\linewidth,keepaspectratio=true]{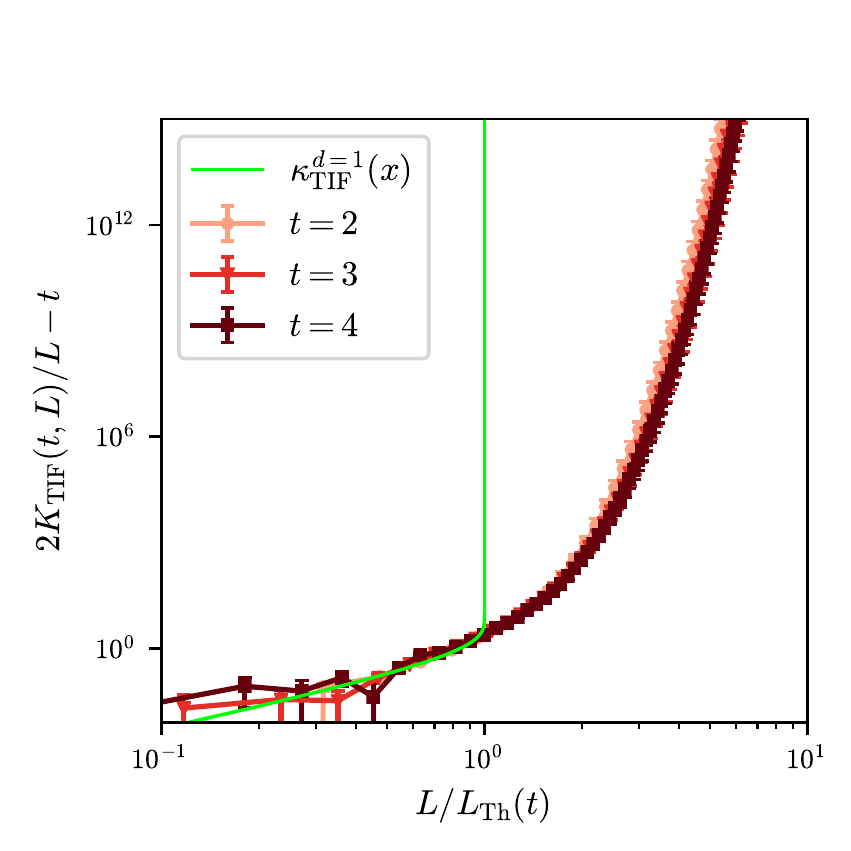}
\end{minipage}
\caption{
 $K(t,L)/\Leff-t$ vs. $L/ \Lth$ for TIF-RPM at $q=3$ (left), TIF-BWM at $q=2$ (middle), and TIF-BWM at $q=3$ (right).
    } \label{fig:caseD_collapse}
\end{figure}

\begin{figure}[ht]
\begin{minipage}[t]{0.31\textwidth}
\includegraphics[width=1.1\linewidth,keepaspectratio=true]{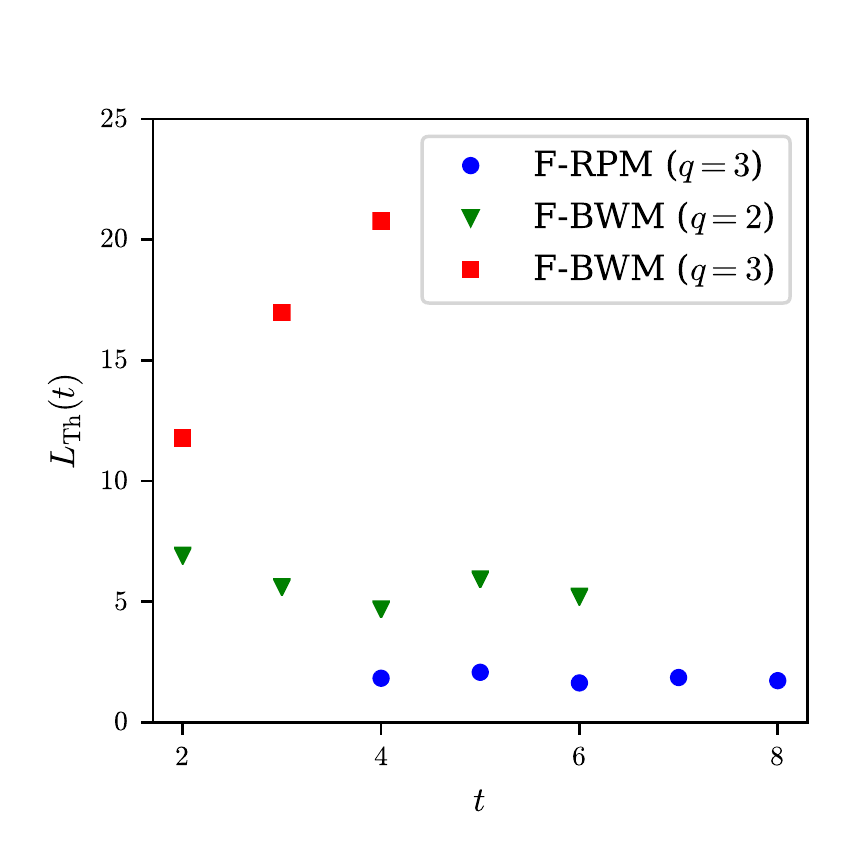}
\end{minipage}
\hspace*{\fill} 
\begin{minipage}[t]{0.31\textwidth}
\includegraphics[width=1.1\linewidth,keepaspectratio=true]{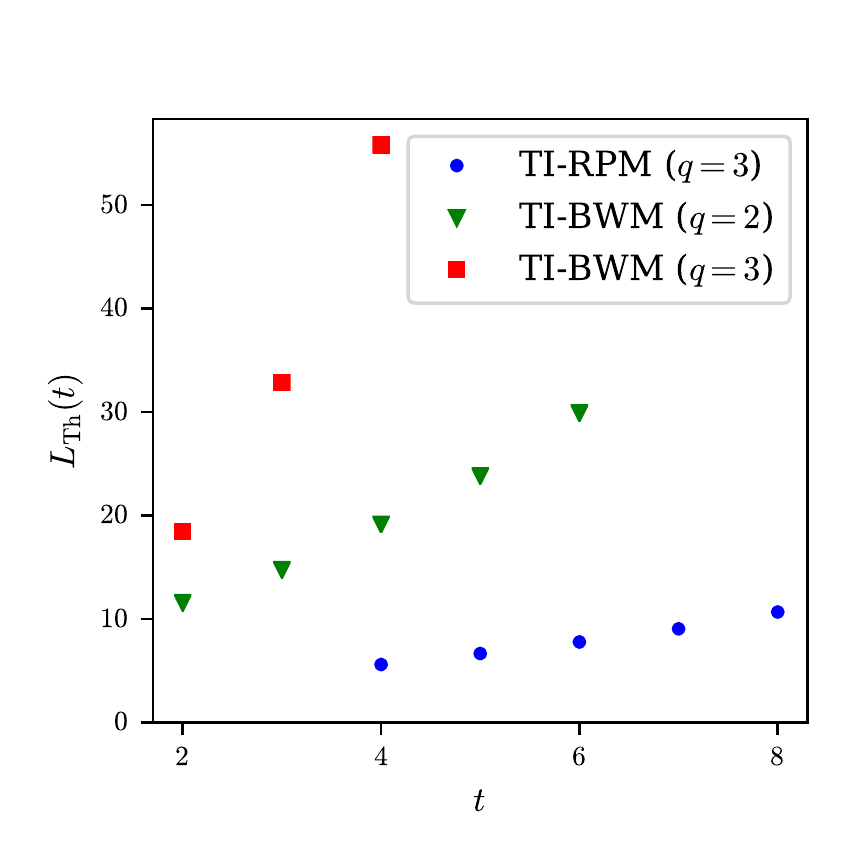}
\end{minipage}
\hspace*{\fill} 
\begin{minipage}[t]{0.31\textwidth}
\includegraphics[width=1.1\linewidth,keepaspectratio=true]{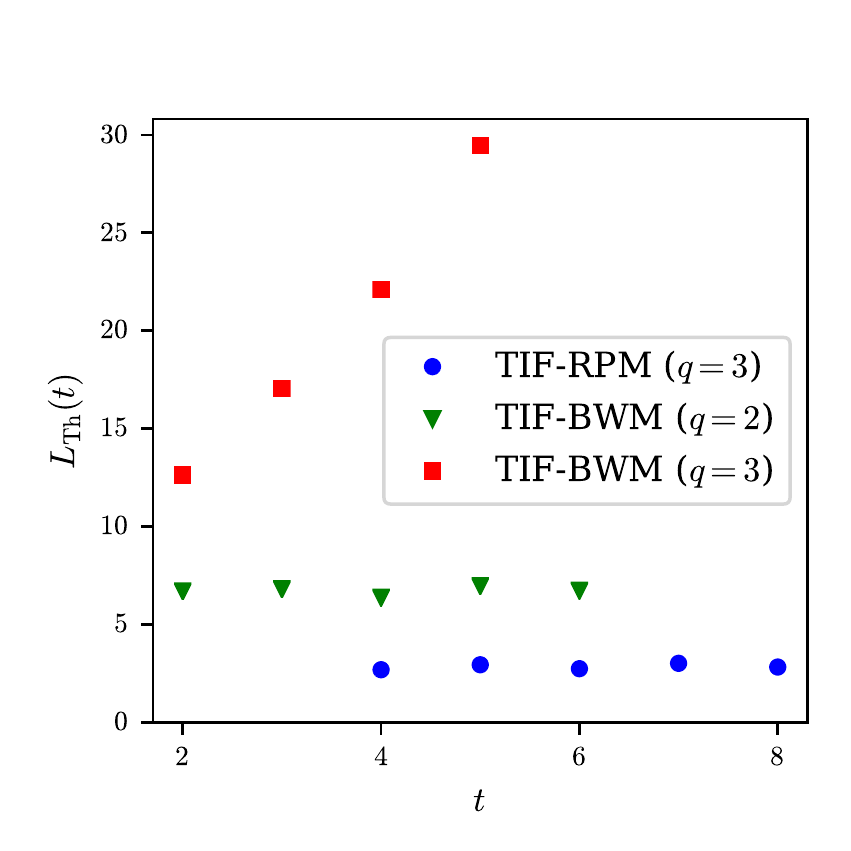}
\end{minipage}
\caption{ 
 $\Lth(t)$ vs. $t$ for Floquet circuits (left),  TI circuits (middle), and TI Floquet circuits (right)
    } \label{fig:L_th_all}
\end{figure}

In this part of the Appendix, we focus on transnational invariant Floquet circuits, and elaborate on the right panel of Figure \ref{fig:collapse}. 
First, we separately show in Figure \ref{fig:caseD_collapse}, $K_{\mathrm{TIF}}/L-t$ vs $L/\Lth(t)$ for RPM with $q=3, \epsilon=2$ (left), BWM with $q=2$ (middle), and BWM with $q=3$. The data is averaged over $8000-15000$ realizations of the dual Floquet operator $V$ for RPM, $7000-1000$ realizations for BWM with $q=2$, and $4000-10000$ realizations for BWM with $q=3$. 
As can be seen in the figure, there aren't enough points for $x=L/\Lth(t)<1$ for RPM with $q=3$ and BWM with $q=2$, in order to see a collapse for $x<1$, hence cannot be compared reliably for $x>1$ either.  
This can be seen as a consequence of small $\Lth(t)$ for those two models (as shown in the right panel of Figure \ref{fig:L_th_all}). 
TIF-RPM with $q=3, \epsilon=2$, in particular, has the additional problem where the extracted $\Lth(t)$ in fact are probing the region beyond the RMT ramp, when $t>\thei \approx q^L/L$. 
Note that this is not an issue for F-RPM, since $\thei = q^L$ for that model. 
These problems are not present in the data for BWM with $q=3$ (right panel of Figure \ref{fig:L_th_all}), because $\Lth(t)$ are sufficiently large for the accessible values of $t$, allowing data points for $x<1$. 
Note that $\Lth(t)$ for $t=5$ is obtained by stochastic sampling of the trace through multiplication with complex random vectors, and is not as accurate as the points for $t \leq 4$.

\begin{figure}[ht]
\begin{minipage}[t]{0.31\textwidth}
\includegraphics[width=1.1\linewidth,keepaspectratio=true]{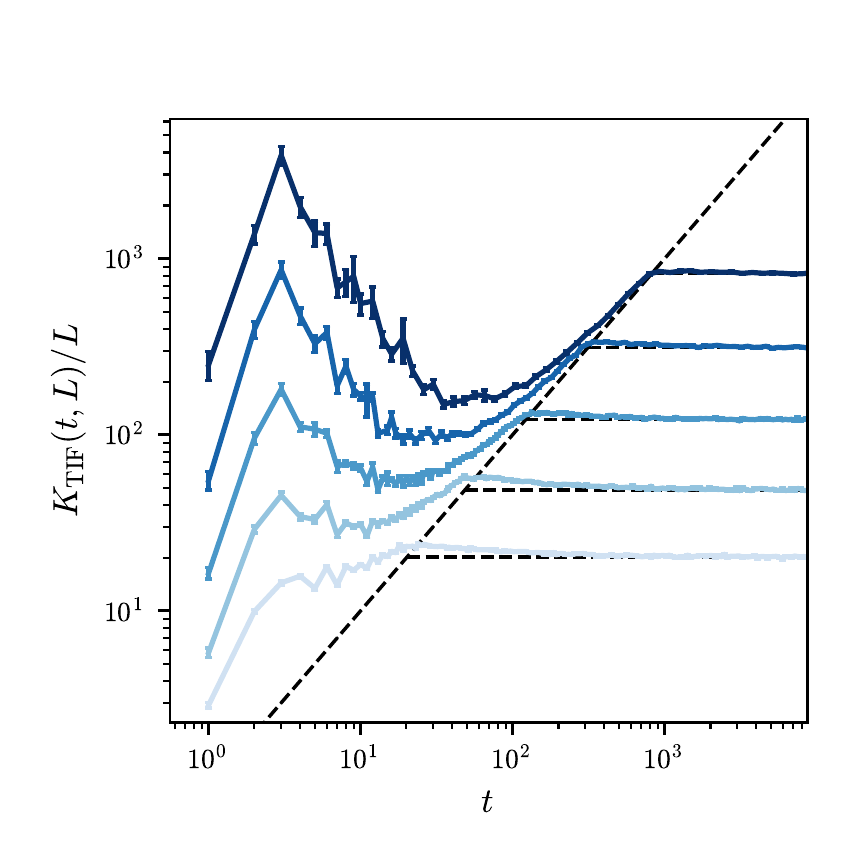}
\end{minipage}
\hspace*{\fill} 
\begin{minipage}[t]{0.31\textwidth}
\includegraphics[width=1.1\linewidth,keepaspectratio=true]{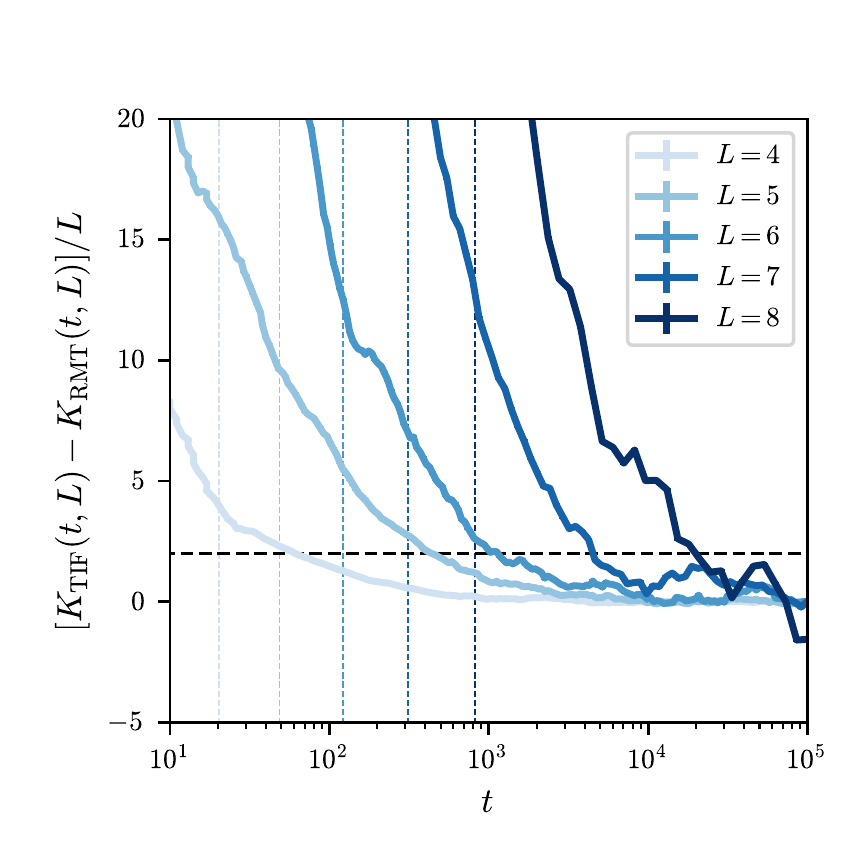}
\end{minipage}
\hspace*{\fill} 
\begin{minipage}[t]{0.31\textwidth}
\includegraphics[width=1.1\linewidth,keepaspectratio=true]{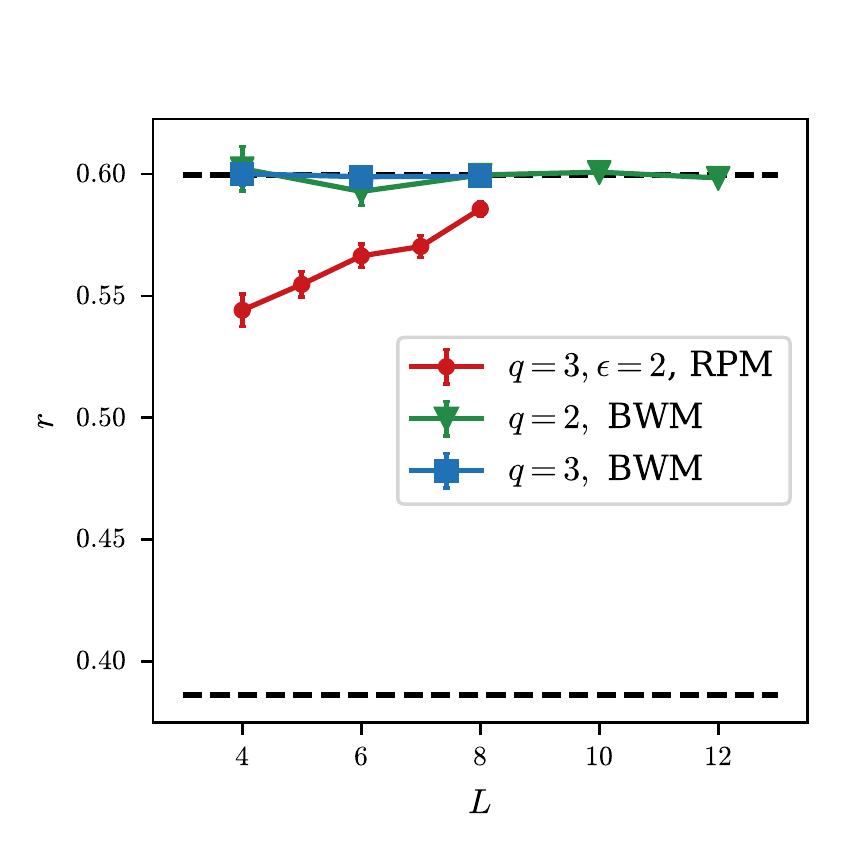}
\end{minipage}
\caption{
 Left : $K_{\mathrm{TIF}}(t,L)/L$ vs. $t$ for TIF-RPM with $q=3, \epsilon=2$, for $L=4,5,6,7,8$ (from light to dark blue), and the dashed lines denote the corresponding RMT behavior defined in Eq. \ref{eq:f_rmt}. Middle : $K_{\mathrm{TIF}}(t,L)/L-\frmt(t,L)/L$ vs. $t$ 
 where the vertical lines correspond to $\thei$ for the $L$ corresponding to the same color. 
 The intersection points with a horizontal line near zero give an estimate of $\tth$ which appear larger than $\thei$. 
 Right: Nearest neighbour level spacing ratio $r$ in the zero momentum sector for the three TIF models studied in this paper.}
 \label{fig:caseD_t_dir}
\end{figure}

To further investigate on the dynamics of TIF-RPM with $q=3,\epsilon=2$, we look at $K_{\mathrm{TIF}}/L$ vs $t$ using time direction simulations, 
and show that $\tth$ is of the order of $\thei$ for accessible $L$. 
The left panel of Figure \ref{fig:caseD_t_dir} plots $K_{\mathrm{TIF}}/L$ vs $t$, averaged over $8000-10000$ realizations of the Floquet operator $W$, along with the RMT behavior $K_{\mathrm{RMT}}$, given by 
\begin{equation}\label{eq:f_rmt}
    \frmt(t,L)=\begin{cases}
      tL & t<\thei\\
      q^L & t\geq \thei
    \end{cases} \;.
\end{equation}
It can be observed in the middle panel that the $K_{\mathrm{TIF}}/L$ approaches $K_{\mathrm{RMT}}/L$ at times larger than $\thei$. The middle panel shows a rolling average of the data from the plot in the left panel. To quantify $\tth$, one could draw a horizontal line near zero and look at the intersection of the difference between $K_{\mathrm{TIF}}/L$ and $K_{\mathrm{RMT}}/L$. A large $\tth$ is consistent with the small $\Lth(t)$ we see in Figure \ref{fig:L_th_all}, and
is also reflected in the ratio of consecutive level spacings $r=$min$(\delta_n,\delta_{n+1})$/max$(\delta_n,\delta_{n+1})$, where $\delta_n=|\phi_n-\phi_{n-1}|$ is the difference between consecutive eigenphases $\{\phi_n \}$, which is plotted in the right panel, averaging over $100$ realizations of the Floquet operator for each model.
The plot shows that the spacing ratio for TIF-RPM hasn't converged to the GUE value for the accessible system sizes, unlike the other two models we have studied.


\subsection{Comparison between scaling forms and finite-$t$, finite-$L$, and infinite-$q$ solutions}
\begin{figure}[ht]
\begin{minipage}[t]{0.31\textwidth}
\includegraphics[width=1.1\linewidth,keepaspectratio=true]{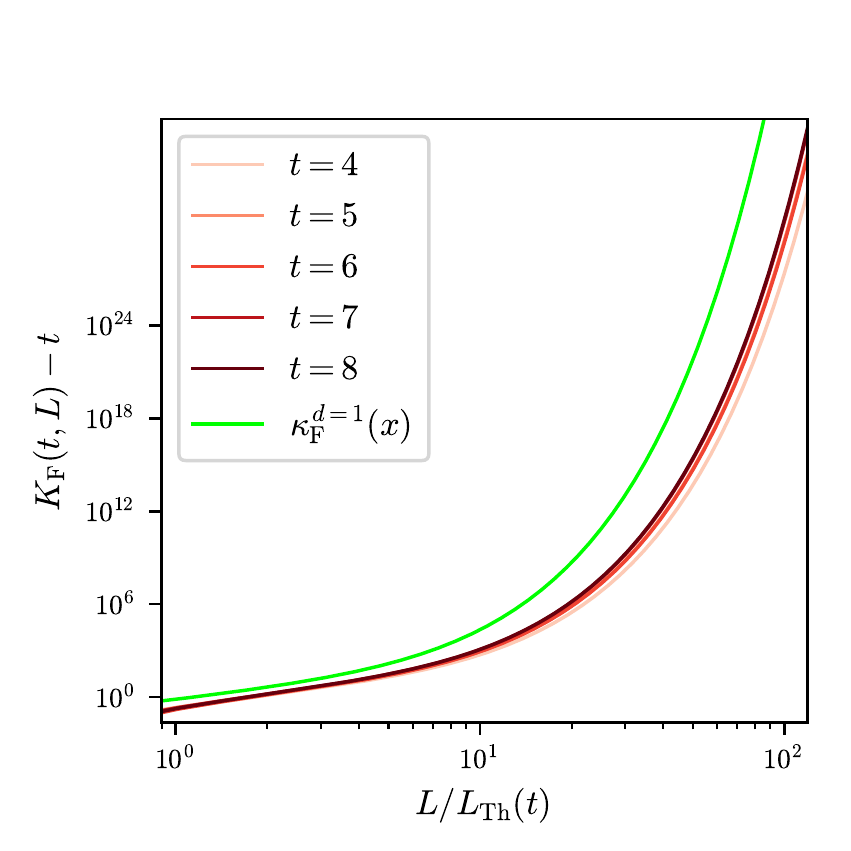}
\end{minipage}
\hspace*{\fill} 
\begin{minipage}[t]{0.31\textwidth}
\includegraphics[width=1.1\linewidth,keepaspectratio=true]{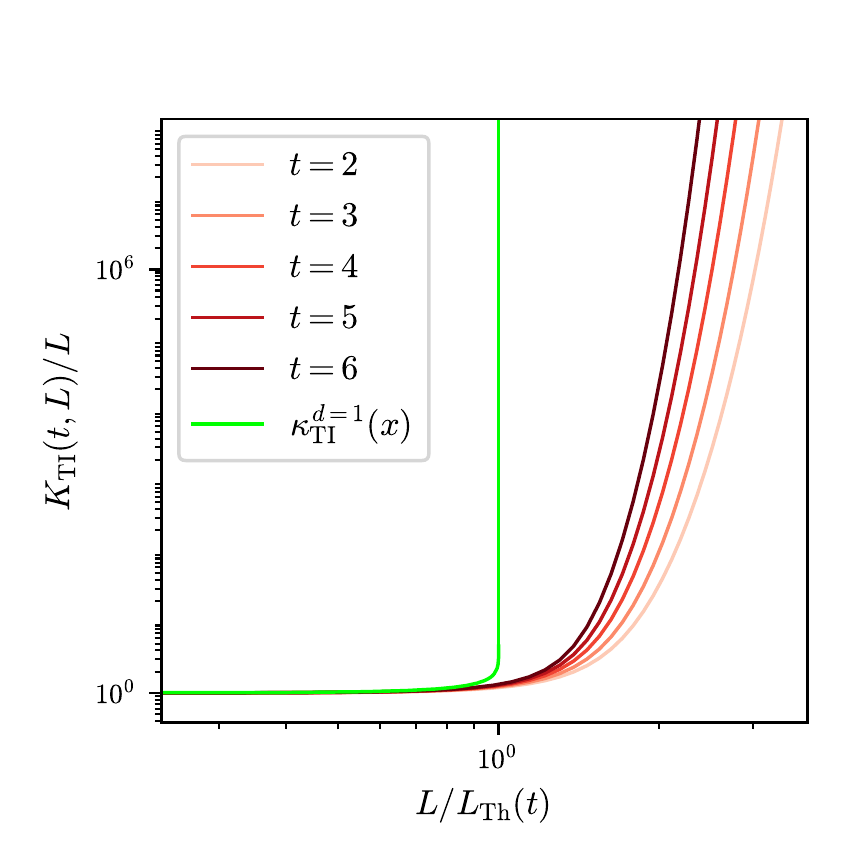}
\end{minipage}
\hspace*{\fill} 
\begin{minipage}[t]{0.31\textwidth}
\includegraphics[width=1.1\linewidth,keepaspectratio=true]{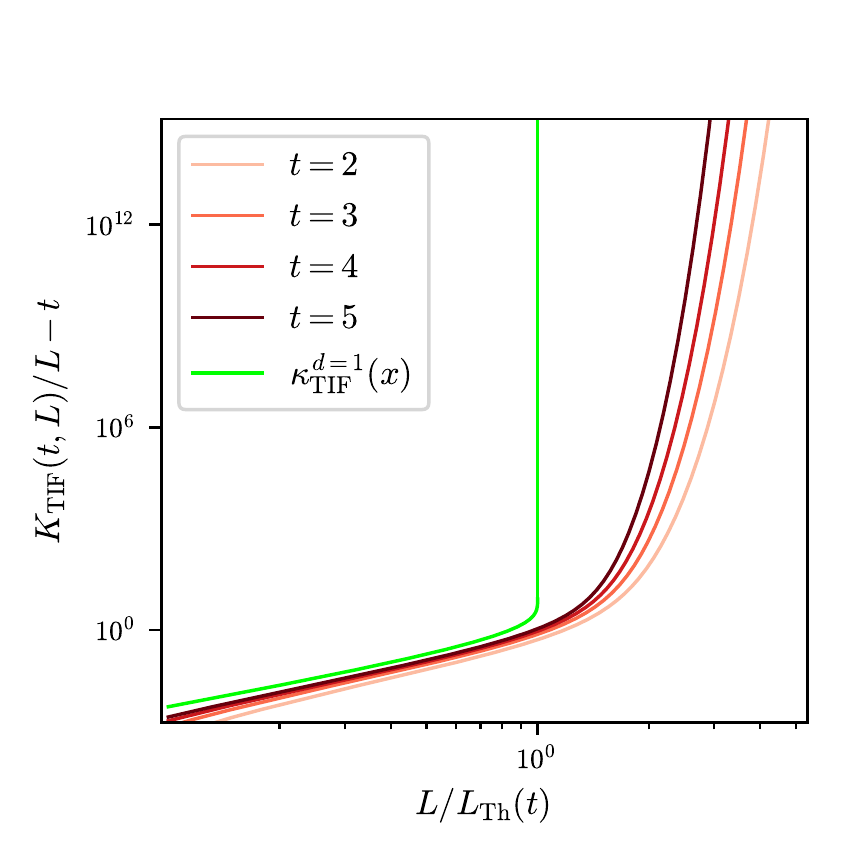}
\end{minipage}
\caption{
For all three plots, the green curves are the infinite-$t$, infinite-$L$, infinite-$q$ scaling functions given in Eq. \ref{eq:kb_scaling_d1}, \ref{eq:kc_scaling}, and \ref{eq:kappaTIFd1} (from left to right respectively). 
 Left: $K_{\mathrm{F}}(t,L)-t$ vs $L/\Lth(t)$ for different $t$ where the red curves denote the finite-$t$, finite-$L$, infinite-$q$ expression for $K_{\mathrm{F}}(t,L)-t$ (Eq. \ref{eq:KFd1lambda}), with $\Lth$ for F-RPM (from Figure \ref{fig:L_th_all}).
 Middle: $K_{\mathrm{TI}}(t,L)/L$ vs $L/\Lth$ different $t$, alongside the finite-$t$, finite-$L$, infinite-$q$ expression for $K_{\mathrm{TI}}(t,L)/L$ (Eq. \ref{eq:sffmultid1}) in reds, with $\Lth$ for TI-BWM with $q=2$. 
 Right: $K_{\mathrm{TIF}}(t,L)/L-t$ vs $L/\Lth$ different $t$, alongside the finite-$t$, finite-$L$, infinite-$q$ expression for $K_{\mathrm{TIF}}(t,L)/L-t$ (Eq. \ref{eq:kd_fin}) in reds, with $\Lth$ for TI-BWM with $q=3$. 
    } \label{fig:fin_L_inf_q_v2}
\end{figure}

As mentioned in the main text, for TI and TIF models, the differences  between the scaling collapse of the finite-$q$ numerics and the infinite-$q$ solution (Fig. \ref{fig:collapse}), can be due to (i) a genuinely different scaling function for finite-$q$, or (ii) a slow convergence in $t$ to the same scaling function in infinite-$q$. 
To investigate the latter possibility, in Figure \ref{fig:fin_L_inf_q_v2}, we compare the finite -$L$, -$t$, infinite-$q$ results for the SFF with the infinie-$q$ scaling functions reached at large $L$ and $t$, for Floquet, TI and TI Floquet models.
%
we find that  finite $t$ corrections decay very slowly for TI systems compared to the Floquet systems.


\end{document}